\documentclass[aps,prd,twocolumn,a4paper,superscriptaddress,10pt,floatfix]{revtex4-2}
\usepackage[american]{babel}
\usepackage{graphicx,hyperref,color,amsmath,amsmath}
\usepackage[dvipsnames]{xcolor}

\begin{document}
\newcommand{\be}{\begin{equation}}
\newcommand{\ee}{\end{equation}}
\newcommand{\bq}{\begin{eqnarray}}
\newcommand{\eq}{\end{eqnarray}}
\newcommand{\bsq}{\begin{subequations}}
\newcommand{\esq}{\end{subequations}}
\newcommand{\bc}{\begin{center}}
\newcommand{\ec}{\end{center}}

\title{Scaling solutions for varying tension strings}
\author{C. S. C. M. Coelho}
\email{up202504528@edu.fc.up.pt}
\affiliation{Centro de Astrof\'{\i}sica da Universidade do Porto, Rua das Estrelas, 4150-762 Porto, Portugal}
\affiliation{Faculdade de Ci\^encias, Universidade do Porto, Rua do Campo Alegre, 4150-007 Porto, Portugal}
\author{A.-L. Y. Gschrey}
\email{An.Gschrey@campus.lmu.de}
\affiliation{Department of Physics, Ludwig-Maximilians-Universität M{\"u}nchen,  Geschwister-Scholl-Platz 1, D-80539 M{\"u}nchen, Germany}
\author{C. J. A. P. Martins}
\email{Carlos.Martins@astro.up.pt}
\affiliation{Centro de Astrof\'{\i}sica da Universidade do Porto, Rua das Estrelas, 4150-762 Porto, Portugal}
\affiliation{Instituto de Astrof\'{\i}sica e Ci\^encias do Espa\c co, Universidade do Porto, Rua das Estrelas, 4150-762 Porto, Portugal}
\date{\today}

\begin{abstract}
We use the Velocity-dependent One Scale Model for topological defect evolution to explore and classify the possible scaling solutions for string networks with time-varying tension, in cosmological and non-cosmological settings and under two different phenomenological assumptions for the behavior of these variations, which rely on different stretching and damping contributions to the string dynamics. We discuss how these assumptions impact the standard scaling solutions, as well as the evolution of the string network density. In addition to simple power-law cosmological epochs solutions, we also discuss the behavior of the network during the radiation-to-matter and matter-to-acceleration transitions. Overall, our results show that for the same amount of tension variation, a change in the stretching length scale tends to have a more significant impact on the network than a change in the damping length.
\end{abstract}
\maketitle

\section{\label{sect1}Introduction}

Topological defects are fossil relics of symmetry-breaking phase transitions, emerging as a result of the Kibble mechanism \cite{Kibble76}. They can therefore form in the early universe \cite{VSbook}, but also in a range of condensed matter systems \cite{Condmat1,Condmat2,Condmat3}. Being non-linear objects, a reliable understanding of their dynamics and consequences requires a careful combination of analytical modeling and numerical analysis. The present work addresses the first of these approaches.

The paradigm underlying rigorous analytic models is to adopt effective descriptions in which the network is treated as a thermodynamical system, in other words, to model the network through suitable averaged (macroscopic) quantities, whose dynamics is described by evolution equations derived, wholly or in part, from the microscopic dynamics---ultimately, for the case of standard cosmic strings, from the Nambu-Goto action. This approach, pioneered by Kibble \cite{Kibble85}, led to the subsequent development of the now canonical Velocity-dependent One Scale (VOS) model \cite{VOS1,VOS2,VOS3,VOSbook}. A self-contained and detailed description of the numerical calibration of this model, both for cosmic strings and domain walls, can be found in \cite{Thesis} and references therein. Various subsequent extensions of the VOS model, e.g. to other types of defects and to those with additional degrees of freedom, have been described in the literature.

In this work, we consider two phenomenological approaches for describing varying tension cosmic strings. The first of these introduces a time dependence in the string tension, $\mu(t)$, at the level of the dynamical VOS macroscopic equations \cite{Yamaguchi1,Yamaguchi2}, while a more recent alternative approach introduces this time dependence in the microscopic equations of motion, and then this is propagated into the VOS equations of motion \cite{Conlon,Revello}. Both of these works restrict their analysis to the better known linear (scale invariant) scaling solution; our present goal is to carry out a more thorough analysis of the impact of the varying tension on all canonical scaling solutions, both in cosmological and non-cosmological contexts. For the same reason we consider both simple power law cosmological epochs and the physically more realistic cases of the radiation to matter and matter to acceleration transitions, which provide additional physical insight on these scaling solutions.

We must start by pointing out that both of the above approaches are purely phenomenological, since they assume that the string tension only depends on a time coordinate (which can be physical or conformal time), without any spatial variation. A full \textit{ab initio} description would rely on additional degrees of freedom on the string worldsheet, e.g. as in the wiggly strings case \cite{Vieira}, whose scaling solutions, under mesoscopic coarse-graining simplifying assumptions, have been reported in \cite{Almeida_2021,Almeida_2022}. Nevertheless, a detailed analysis of the scaling solutions in these $\mu(t)$ cases is warranted for several reasons. One of these is that, although the two approaches are superficially unrelated, in a modeling sense, and especially in the cosmological context, the two correspond to opposite assumptions on the physical impact of the varying tension. Specifically, in the microscopic approach we are effectively rescaling the damping term in the network's dynamical equations without affecting the stretching term, while in the macroscopic case we are doing exactly the opposite.

The structure of this work is as follows. We start in Sect. \ref{sect2} with a brief overview of the VOS model and its proposed varying tension extensions, following which the non-cosmological and cosmological power law scaling solutions are reported in Sects. \ref{sect3} and \ref{sect4} respectively. Section \ref{sect5} is a slight digression on the case of power-law contracting universes. In Sect. \ref{sect6} we discuss the behavior of these networks in realistic cosmological settings, specifically during the transitions from radiation to matter domination, and from matter domination to acceleration (assuming, in the latter case, a flat $\Lambda$CDM model). In these transitions the networks are not exactly scaling, but their evolution can be approximately studied as a perturbation on relevant scaling solutions. Finally, Sect. \ref{sect7} contains our conclusions. Unless otherwise stated, we work in units with the speed of light $c=1$.

\section{\label{sect2}The VOS model and its varying-tension extensions}

The VOS model is the canonical quantitative model for the evolution of topological defect networks, in cosmological and other contexts, and has been subject to extensive numerical validations. In what follows we describe its salient conceptual features and define the relevant quantities, referring the reader to the original works \cite{VOS1,VOS2,VOS3} for further details, including the derivations of the evolution equations.

There are two key conceptual assumptions underlying the model. The first is localization: the assumption that we can treat strings as one-dimensional line-like objects. This should be adequate for local strings, while for global strings, which have long-range interactions, it is more questionable but can also be justified through comparison with field theory numerical simulations \cite{Moore}. The second assumption is that the microscopic equations of motion can be averaged to yield macroscopic (thermodynamical) counterparts. With the further assumption that on large scales the long strings network is a random walk, the two dynamical variables in the VOS model are the root mean square velocity, $v$, and a characteristic length scale, $L$ (alternatively interpretable, in the VOS context, as a correlation length, inter-string distance, or string curvature radius), related to the string density through
\be
\rho=\frac{\mu}{L^2}\,,
\ee
where $\mu$ is the string mass per unit length. Note that this relation implies that even if the scaling behavior of $L$ is unchanged with respect to the standard case, changes in $\mu$ will by themselves impact the ratio of string and background densities. Conversely, changes in the scaling behavior of $L$ can be offset by the changed behavior of $\mu$ leading to an unchanged behavior of the density. Examples of both situations will be seen in the following sections.

With those assumptions, one obtains the following evolution equations for the two dynamical variables
\begin{subequations}
\bq
2\frac{dL}{dt} &=& \frac{L}{\ell_s}+\frac{L}{\ell_d}v^2+{\tilde c}v\\
\frac{dv}{dt} &=& (1-v^2)\left[\frac{k(v)}{L} -\frac{v}{\ell_d}\right]\,,
\eq
\end{subequations}
where we have defined the stretching length and damping length, respectively, by
\begin{subequations}
\bq
\frac{1}{\ell_s}&=&2H\\
\frac{1}{\ell_d}&=&2H+\frac{1}{\ell_f}\,,
\eq
\end{subequations}
where $H=(1/a)(da/dt)$ is the Hubble parameter and $\ell_f$ is a friction length scale, nominally due to particle scattering. Specifically, for a local cosmic string network the dominant effect comes from Aharonov-Bohm scattering, and we can write \cite{Vilenkin}
\be\label{particlescattering}
\ell_f=\frac{\mu}{\theta T^3}\,,
\ee
where $T$ is temperature and $\theta$ is a numerical factor related to the number of particle species interacting with the strings, which is expected to be of order unity for strings forming in minimal Grand Unified Theory scenarios, but can be larger in extensions thereof. Note that the damping length explicitly impacts both dynamical quantities (naturally in a velocity-dependent way), while the stretching length only does so for the characteristic length scale.

In its simplest version discussed herein, the model has two free parameters, the loop chopping efficiency ${\tilde c}$ (not to be confused with the speed of light), and the momentum parameter $k(v)$, which is a function of velocity \cite{VOS3}, specifically
\be\label{defk}
k(v)=\frac{2\sqrt{2}}{\pi}(1-v^2)(1+2\sqrt{2}v^3)\frac{1-8v^6}{1+8v^6}\,.
\ee
These parameters can be calibrated using both Nambu-Goto and field theory simulations. This procedure and its quantitative outcome are described in detail, for an extended six-parameter version of the model, in \cite{Correia:2019bdl,Correia:2021tok}. This extended version of the model is not necessary for our present purposes, and our results will not depend on this choice. The reason is that these changes, which impact the energy loss terms and $k(v)$, are themselves velocity-dependent: for low or decaying velocities the extra parameters are not relevant, while for constant velocities they can be accounted for by renormalizing $\tilde c$ and $k(v)$, both of which are constants in this limit.

This being said, it is now straightforward to describe the two phenomenological approaches to time-dependent tension strings previously suggested in the literature. When the time dependence is introduced macroscopically \cite{Yamaguchi1,Yamaguchi2}, the outcome is that the logarithmic drift of the string tension is added to the Hubble parameter contribution in the stretching term
\be\label{rescales}
\frac{1}{\ell_s}=2H+\frac{1}{\mu}\frac{d\mu}{dt}\,,
\ee
while the damping length is unchanged. Therefore, only the length equation is explicitly changed---although the velocity evolution is still impacted since the two equations are coupled. Conversely, if the phenomenological varying tension is introduced microscopically \cite{Conlon,Revello}, the logarithmic drift of the string tension emerges as an additional contribution to the damping length
\be\label{rescaled}
\frac{1}{\ell_d}=2H+\frac{1}{\ell_f}+\frac{1}{\mu}\frac{d\mu}{dt}\,,
\ee
while the stretching length is unchanged. In this case, both dynamical equations are explicitly changed.

In passing, it is interesting to contrast the above with the case of wiggly cosmic strings \cite{Vieira}. There, the relevant additional parameter is the renormalized string mass per unit length, herein denoted $\mu_{wig}$ (whose value is unity for plain Nambu-Goto strings), and the stretching and damping lengths both depend on it, but in different ways. Indeed, in this case we can write
\be
\frac{1}{\ell_{s,wig}}=2H+\left(1-\frac{1}{\mu^2_{wig}}\right)H\,
\ee
\be
\frac{1}{\ell_{d,wig}}=2H-\left(1-\frac{1}{\mu^2_{wig}}\right)H+\frac{1}{\ell_f}\,,
\ee
with the caveat that this stretching length applies to the characteristic inter-string distance $L$ (a measure of the network's density). For wiggly strings the network's correlation length differs from it, and effectively has a different stretching length which depends on the scale at which this renormalized mass is defined; see \cite{Vieira} for a detailed discussion of this point. In any case, the interesting point is that in this case the stretching and damping lengths are changed in exactly opposite ways.

Returning to the present phenomenological approach, in both the macroscopic and microscopic cases, if both the scale factor and the string tension have power law time dependencies, one might interpret the varying tension as rescaling the expansion rate, with this rescaling impacting either the stretching or damping terms, but not both simultaneously. Assuming the more plausible case of a decreasing tension, the effect is therefore to decrease the effective expansion rate. For a sufficiently fast decrease we could naively envisage such a scenario as analogous to a contracting universe, a scenario mentioned in passing in \cite{Revello} which we will further explore.

For completeness, it is also of interest to consider the VOS model evolution of a specific string segment of length $\ell$, e.g. a loop. In this case the evolution equation is
\be\label{loopeqn}
\frac{d\ell}{dt}=\frac{\ell}{\ell_s}-\frac{\ell}{\ell_d}v^2+\left(\frac{d\ell}{dt}\right)_{rad}\,,
\ee
where in this case $1/\ell_s=H$ and the last term accounts for radiation losses. For the simplest plain strings these losses would be through gravitational radiation, but that will be subdominant for more realistic defects which include additional degrees of freedom on the string worldsheet, e.g. for superconducting strings the dominant energy loss mechanism will be electromagnetic radiation. In any case the stretching or damping lengths, depending on the case, would also be changed as described above.

In what follows we start by assuming power-law dependencies both for the scale factor and the string tension,
\begin{subequations}
\bq
a(t)&\propto&t^\lambda\\
\mu(t)&\propto&t^\delta\,,
\eq
\end{subequations}
and look for power-law scaling solutions for the characteristic length scale and rms velocity, in cosmological and other settings. We are primarily interested in expanding universes with $0<\lambda<1$, and, at least in cosmological settings, in the case of decaying tensions ($\delta<0$), but in the latter part of the work we will also discuss other contexts and alternative assumptions for the tension evolution.

\section{\label{sect3}Non-cosmological scaling solutions}

While our main interest is in cosmological solutions, in this section we briefly consider two non-cosmological settings: condensed matter and Minkowski spacetime. The former is self-evidently experimentally relevant, while the latter is useful since it can be explored through numerical simulations and provides a limiting case of vanishing cosmological expansion. Additionally, both of these settings will provide early insights which will be relevant for the cosmological context.

\subsection{Condensed matter}

In the condensed matter case we have $H=0$ and $\ell_f=const$. For easier comparison with the condensed matter literature one can alternatively define $\Gamma=c \ell_f$. (For convenience, in this condensed matter setting the speed of light is included explicitly, rather than set to $c=1$ as in the rest of the work.) In the standard case the evolution equations are therefore
\begin{subequations}
\bq
2\frac{dL}{dt} &=& \frac{Lv^2}{\Gamma}+{\tilde c}v\\
\frac{dv}{dt} &=& (c^2-v^2)\left(\frac{k}{L} -\frac{v}{\Gamma} \right)\,,
\eq
\end{subequations}
and the physical scaling solution, well known in condensed matter systems, is
\begin{subequations}\label{condmatstd}
\bq
L &=& \sqrt{k(k+{\tilde c})}\, (\Gamma t)^{1/2}\\
v &=& \sqrt{\frac{k}{k+{\tilde c}}}\,\left(\frac{\Gamma}{t}\right) ^{1/2}\,,
\eq
\end{subequations}

In this case the impact of the varying tension extension depends on how it is modeled. If one still  enforces $\ell_f=const$, then the standard solution is unchanged in the microscopic case (since by assumption the added effective damping term decays with time), while in the macroscopic case the time dependence of the scaling law is not affected, but its normalizations are changed to
\begin{subequations}
\bq
L &=& \sqrt{\frac{k(k+{\tilde c})}{1-\delta}}\, (\Gamma t)^{1/2}\\
v &=& \sqrt{\frac{k(1-\delta)}{k+{\tilde c}}}\,\left(\frac{\Gamma}{t}\right) ^{1/2}\,;
\eq
\end{subequations}
a decreasing tension increases the network's rms velocity and decreases the correlation length. Moreover, in either case the network density, $\rho=\mu/L^2\propto\mu/t$, will be affected by a varying tension.

However, a physically more realistic condensed matter setting corresponds to the assumption of constant temperature, in which case the friction length becomes time dependent if the string tension does. In this case, in the macroscopic description, there is a generalized version of the solution of Eq. (\ref{condmatstd}), which can be written
\begin{subequations}\label{ctemp1}
\bq
L &=& \sqrt{k(k+{\tilde c})}\, ({\tilde \mu}\Gamma t)^{1/2}\\
v &=& \sqrt{\frac{k}{k+{\tilde c}}}\,\left({\tilde \mu}\frac{\Gamma}{t}\right) ^{1/2}\\
|\delta| &<1&\,,
\eq
\end{subequations}
where for convenience, and by analogy with the wiggly strings case \cite{Vieira},  we have defined the dimensionless parameter ${\tilde\mu}=\mu(t)/\mu_0$ as the time-dependent part of the string tension, i.e. the excess (or deficit) tension as compared to its baseline value. (Conversely, $\Gamma$ is still a constant, without this time dependence.) In this case a varying tension impacts both scaling laws: a decreasing tension makes the characteristic length grow more slowly and the velocities decay faster. Interestingly the network density is unaffected. Note that the exponent $\delta$ is bounded by two physical limits: for $\delta=-1$ one would have a constant correlation length (with decaying velocities), while for the causal limit $\delta=1$ one would have a linear-type scaling solution (with constant velocities). A solution with $\delta>1$, with both the velocity and correlation length growing, and the latter doing so faster than linearly, would evidently be a transient one, and moreover would be akin to the cosmological Kibble regime \cite{Kibble85}.

In the microscopic case, an almost identical extension obtains, with the normalization factors also being impacted,
\begin{subequations}\label{ctemp2}
\bq
L &=& \sqrt{\frac{k(k+{\tilde c})}{1+\delta}}\, ({\tilde \mu}\Gamma t)^{1/2}\\
v &=& \sqrt{\frac{k(1+\delta)}{k+{\tilde c}}}\,\left({\tilde \mu}\frac{\Gamma}{t}\right) ^{1/2}\\
|\delta| &<1&\,,
\eq
\end{subequations}
and the same physical limits apply for the exponent $\delta$.

We note that in the macroscopic case, the varying tension effectively introduces a stretching term (which would not otherwise exist in condensed matter settings), while in the microscopic case the damping term is changed. However, if one has $|\delta|<1$ the effect of this additional damping term is always subdominant with respect to the changing tension in the ordinary friction term.

\subsection{Minkowski spacetime}

An even simpler case is that of Minkowski spacetime, with $H=0$ and $l_f=\infty$, in which case the standard scaling law is
\begin{subequations}\label{mink}
\bq
L&=&\frac{1}{2}{\tilde c}v_0t\\
v&=&v_0\,.
\eq
\end{subequations}
Mathematically any value of $v_0$ is in principle allowed (including the limiting case $v_0=1$, which is clearly unphysical), but relying on the insight that one should have $k(v)=0$ in this case \cite{VOS3,VOSbook}, we naturally expect that $v_0=1/\sqrt{2}$ is the physically preferred solution, at least in the standard case of constant-tension strings.

In this case, if one interprets the varying tension as being akin to a stretching term the above solution generalizes to
\begin{subequations}
\bq
L&=&\frac{{\tilde c}v_0}{2-\delta}t\\
v&=&v_0\,.
\eq
\end{subequations}
subject to the obvious $\delta<2$ constraint. On the other hand, if one interprets the varying tension as being akin to a damping term the above solution generalizes to
\begin{subequations}
\bq
L&=&\frac{{\tilde c}v_0}{2(1-\delta)}\frac{t}{\tilde\mu}\\
v&=&\frac{v_0}{\tilde\mu}\,.
\eq
\end{subequations}
Nominally, this is subject to the more stringent $0\le\delta<1$ constraint, in which case the velocity decreases and the correlation length grows more slowly than linearly. Both of these are to be expected, since phenomenologically one effectively has a positive damping term. As a matter of fact, mathematically, a value of $-1<\delta<0$ also allows this solution, provided $v_0<<1$. This last feature is not expected to be a physically natural value for the velocity in Minkoswki space, but we note that this solution would again be a transient Kibble-like one. Last but not least, it is worthy of note that in both of these scenarios, it is still the case that $k=0$, as expected.

Thus if the varying tension is envisaged as an effective stretching the scaling laws themselves are unaffected (and only the normalization of the characteristic length is impacted), while if it is envisaged as an effective damping term both scaling laws are affected. Moreover, in Minkowski space the Kibble-like solution occurs for a decaying tension, while in the previously discussed condensed matter case it occurs for a growing tension. This difference is due to the fact that in the former case of this specific solution the varying tension is envisaged as damping (so a negative sign damping term is needed if velocities are to increase), while in the latter it was envisaged as a stretching term.

\section{\label{sect4}Standard Cosmological scaling solutions}

We now shift our attention to the cosmological context. In the standard constant tension case there are three main solutions \cite{VOS1,VOS2}: the transient (friction-dominated) stretching and Kibble solutions apply at early times, while the best known linear (scale-invariant) scaling solution is the true late-time attractor. The transition from one regime to the other will depend on the epoch of network formation, or equivalently on the string symmetry breaking scale: for heavier strings, the friction-domination epoch is shorter and linear scaling is reached earlier.

\subsection{Early times}

Starting with the early-time friction-dominated cases, we recall from Eq. (\ref{particlescattering}) that 
\be
\ell_f= \frac{\mu}{\beta T^3}\propto {\tilde\mu} t^{3\lambda}\propto t^{3\lambda+\delta}\,,
\ee
which shows that for $\delta>0$ the friction length scale grows faster than in the canonical case, and therefore the transient friction-dominated phase will be shorter, while for $\delta<0$ it will be longer (delaying the approach to the linear scaling attractor).

The most relevant cosmological epoch is the radiation era, since the damping due to particle scattering only dominates at early times, gradually becoming subdominant with respect to the expansion term. For reference, in the radiation era $\ell_f\propto {\tilde\mu} t^{3/2}$, while in the matter era $\ell_f\propto {\tilde\mu} t^{2}$. It's also worthy of note, following Kibble \cite{Kibble85}, that in damped regimes the velocity is expected to change slowly, so $v\propto l_f/L$. This still applies in the macroscopic case (though note that $\mu$ is itself time-dependent), and the same is true for the microscopic one provided the running of the tension is sufficiently slow. Mathematically one could envisage an opposite limit of fast running tension, for which $v\propto k\mu/(L{\dot\mu})$, but this would require a large positive $\delta$ and is therefore not expected to be a physically realistic scenario.

The transient stretching regime occurs when the density and velocity are sufficiently low to make long string intercommutations and loop production negligible, in which case the scaling solution is
\begin{subequations}
\bq
L &\propto& a\\
v &\propto&  \frac{l_f}{a} \propto a^2\,.
\eq
\end{subequations}
This corresponds to $L\propto t^{1/2}$ and $v\propto t$ in the radiation era, and to $L\propto t^{2/3}$ and $v\propto t^{4/3}$ in the matter era. Thus the string density dilutes more slowly than the background density, and the network velocity increases as the strings' characteristic length decreases when compared to the size of the horizon. In the macroscopic phenomenological case, this solution generalizes to
\begin{subequations}
\bq
L &\propto& \sqrt{\tilde\mu}a\\
v &\propto& \sqrt{\tilde\mu} a^2\\
-2\lambda <&\delta&<2(1-\lambda)\,.
\eq
\end{subequations}
The behavior of $L$ is evident, given that in this approach the varying tension impacts the stretching term. Note that in the limit $\lambda\to0$ this solution only exists for non-negative values of $\delta$, while in the limit $\lambda\to1$ we have the opposite requirement. A decaying tension leads to the slower growth of both quantities. Note that the lower limit would lead to constant $L$ with growing velocity, while the upper limit comes from the physical need to avoid superluminal growth, which is physically unacceptable in a stretching solution. On the other hand, the evolution of the density itself is never affected. Moving on to the microscopic case, the stretching solution becomes
\begin{subequations}
\bq
L &\propto& a\\
v &\propto&\tilde\mu a^2\,.
\eq
\end{subequations}
In this case it is equally clear that the behavior of $L$ should not be impacted. Still, a decaying tension leads to a slower growth of the velocities. For $\mu\propto a^{-2}$ we will have a constant velocity and $\rho=\mu/L^2\propto a^{-4}$, meaning that the network density would be a constant fraction of the (radiation) background one. The analogous behavior would ensue in the matter era for a string tension decaying as $\mu\propto a^{-1}$. For an even faster tension decay the velocity would decrease, making this solution an attractor (rather than a transient solution). In this case the string density would grow relative to that of the background, and ultimately the string network would dominate the universe's energy density \cite{Dominated1,Dominated2}, in direct conflict with observations. This fast decay scenario is therefore expected to be physically unrealistic.

Even if the friction domination persists, velocities and energy losses will, under ordinary circumstances, eventually become non-negligible. At this point the physical conditions leading to the stretching regime no longer apply, and we enter the Kibble regime
\begin{subequations}
\bq
L &\propto& \left[\frac{k(k+{\tilde c})}{1+\lambda}\ell_f t \right]^{1/2}\\
v &\propto&  \left[\frac{k(1+\lambda)}{k+{\tilde c}}\frac{\ell_f}{t}\right]^{1/2}\,;
\eq
\end{subequations}
thus in the radiation era we have $L\propto t^{5/4}$ and $v\propto t^{1/4}$, while in the matter era we would have $L\propto t^{3/2}$ and $v\propto t^{1/2}$. Physically we have a high-density network (with a correspondingly small correlation length) rapidly losing energy through loop production, and diluting faster than the background density. The network's average velocity still increases, although more slowly than in the stretching regime since fast-moving regions have an enhanced probability of being incorporated into loops which leave the network.

In this branch of the solutions, and for the macroscopic case, the Kibble scaling solution becomes
\begin{subequations}
\bq
L &\propto& \left[\frac{k(k+{\tilde c})}{1+\lambda}{\tilde\mu}\ell_f t \right]^{1/2}\\
v &\propto&  \left[\frac{k(1+\lambda)}{k+{\tilde c}}\frac{{\tilde\mu}\ell_f}{t}\right]^{1/2}\,;
\eq
\end{subequations}
as in the stretching regime for this case, a decaying tension leads to the slower growth of both quantities; the two behaviors are of course related: with lower velocities, loop production becomes less efficient, and so does the network's ability to lose energy. Still, the behavior of the network density is again unaffected, since the string tension is also changing. An almost identical solution also applies for the microscopic case
\begin{subequations}
\bq
L &\propto& \left[\frac{k(k+{\tilde c})}{1+\lambda+\delta}{\tilde\mu}\ell_f t \right]^{1/2}\\
v &\propto&  \left[\frac{k(1+\lambda+\delta)}{k+{\tilde c}}\frac{{\tilde\mu}\ell_f}{t}\right]^{1/2}\,;
\eq
\end{subequations}
For these solutions to retain their Kibble-like character, one must have $\delta>1-3\lambda$; this corresponds to $\delta>-1/2$ for the most relevant case of the radiation era. In the limiting case of $\delta=1-3\lambda$ one would have a linear scaling regime $L\propto t$ and $v\propto const$.

It is worthy of note that mathematically these Kibble scaling solutions are generalizations of the constant temperature condensed matter solutions discussed in the previous section, cf. Eqs. (\ref{ctemp1}--\ref{ctemp2}), although their physical interpretation and the bounds on the behavior of the tension are somewhat different in the two contexts.

\subsection{Late times}

We now consider the late-time behavior, applicable once friction due to particle scattering becomes negligible. It is enlightening to start by the particular case ${\tilde c}=0$; here the physical solution is
\begin{subequations}
\bq
L&=&\frac{k}{2\sqrt{\lambda(1-\lambda)}}t\\
v&=&\sqrt{\frac{1-\lambda}{\lambda}}\,;
\eq
\end{subequations}
with a velocity-related restriction: with the weak constraint $v_0^2\le1$ we must have $\lambda\ge1/2$, while if the constraint is the physically more realistic  $v_0^2\le1/2$ we have $\lambda\ge2/3$. This confirms the point, known since the first generation of Nambu-Goto cosmic string simulations \cite{Bennett:1989yp,Allen:1990tv}, that a linear scaling solution can occur in the matter-dominated era even in the absence of loop production. This solution straightforwardly generalizes for varying tensions. In the macroscopic case, it becomes
\begin{subequations}
\bq
L&=&\frac{k}{2\sqrt{\lambda(1-\lambda-\delta/2)}}t\\
v&=&\sqrt{\frac{1-\lambda-\delta/2}{\lambda}}\,,
\eq
\end{subequations}
and the previous weak and physical constraints become $\lambda\ge(1-\delta/2)/2$ and $\lambda\ge(2-\delta)/3$ respectively, whence it follows that for $\delta\ge1/2$ a linear scaling solution without loop production would also become possible in the radiation era. Instead, in the microscopic case the scaling solution has the form
\begin{subequations}
\bq
L&=&\frac{k}{2\sqrt{(\lambda+\delta/2)(1-\lambda)}}t\\
v&=&\sqrt{\frac{1-\lambda}{\lambda+\delta/2}}\,;
\eq
\end{subequations}
here the weaker constraint remains the same while the physical one becomes $\lambda\ge(2-\delta/2)/3$, and therefore a radiation era linear scaling solution without loop production has the stronger requirement $\delta\ge1$.

\begin{figure*}
    \centering
    \includegraphics[width=0.49\textwidth,keepaspectratio]{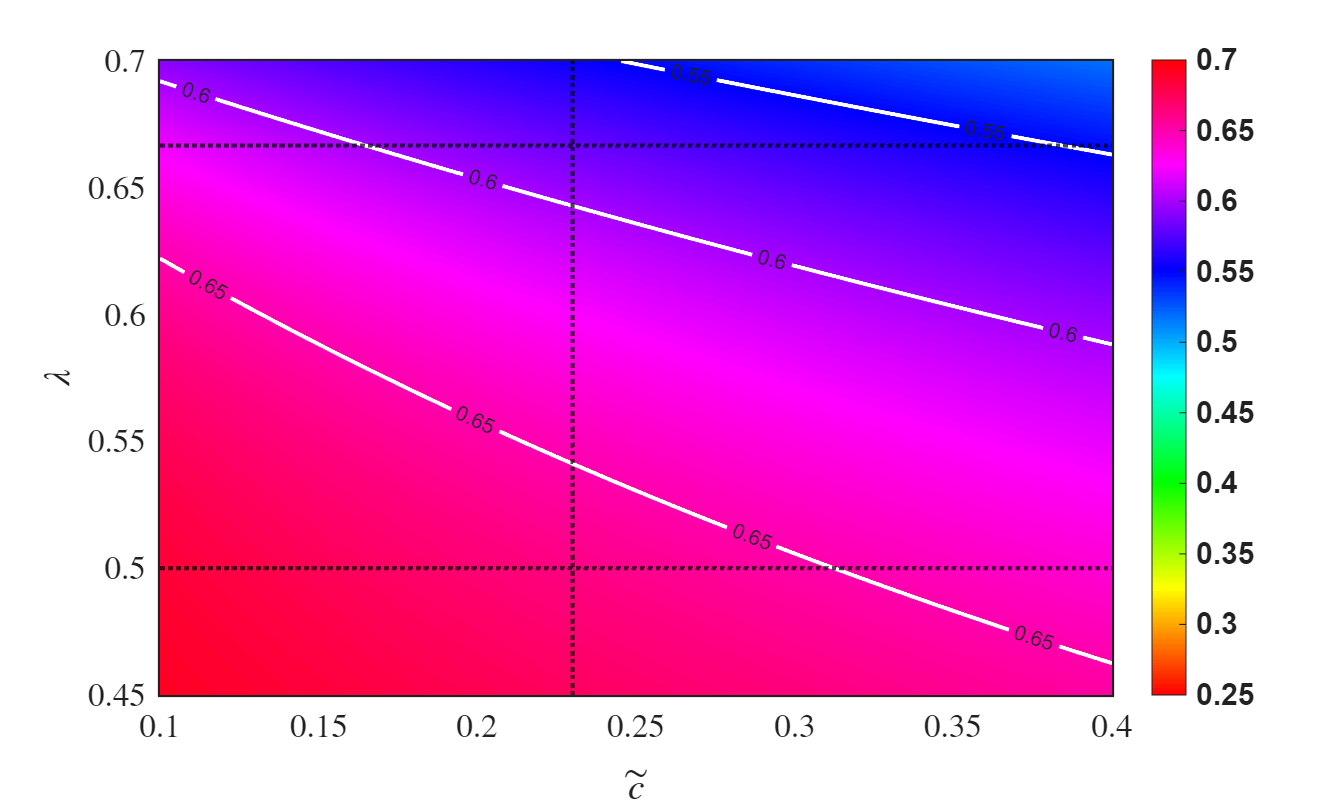}
    \includegraphics[width=0.49\textwidth,keepaspectratio]{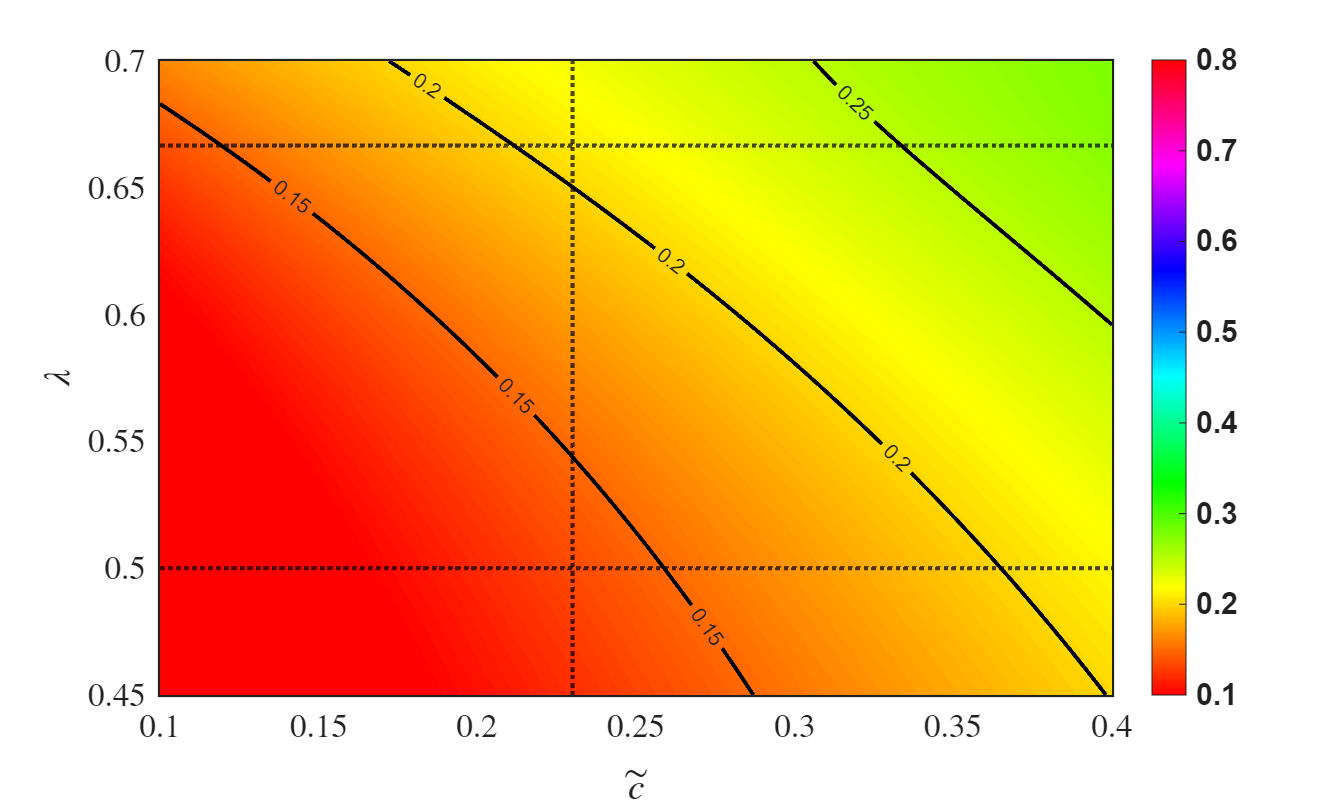}
    \caption{The linear scaling values of $v$ and $\epsilon$ (left and right panels respectively) for constant-tension strings, for relevant values of the expansion rate $\lambda$ and the loop chopping efficiency $\tilde c$. The two horizontal dotted lines identify the radiation end matter eras ($\lambda=1/2$ and $\lambda=2/3$ respectively), while the vertical dotted line identifies the numerically preferred value $\tilde c=0.23$. Some particular contour lines are also drawn for visual convenience.}
    \label{fig1}
\end{figure*}

The main attractor solution for power-law scale factors is the well-known linear scaling solution
\begin{subequations}\label{linear}
\bq
L&=&\frac{1}{2}\sqrt{\frac{k(k+{\tilde c})}{\lambda (1-\lambda)}}\, t\\
v&=&\sqrt{\frac{1-\lambda}{\lambda} \frac{k}{k+{\tilde c}}}\,.
\eq
\end{subequations}
For illustration, Fig. \ref{fig1} shows the scaling values of $v$ and $\epsilon=L/d_H$, where $d_H=t/(1-\lambda)$ is the Hubble radius, for relevant values of the parameters $\lambda$ and $\tilde c$, and using $k(v)$ given by Eq. (\ref{defk}).

Once more, these straightforwardly generalize, in the macroscopic case, to
\begin{subequations}
\bq
L&=&\frac{1}{2}\sqrt{\frac{k(k+{\tilde c})}{\lambda (1-\lambda-\delta/2)}}\, t\\
v&=&\sqrt{\frac{1-\lambda-\delta/2}{\lambda} \frac{k}{k+{\tilde c}}}\,.
\eq
\end{subequations}
while in the microscopic one we have
\begin{subequations}
\bq
L&=&\frac{1}{2}\sqrt{\frac{k(k+{\tilde c})}{(\lambda+\delta/2) (1-\lambda)}}\, t\\
v&=&\sqrt{\frac{1-\lambda}{\lambda+\delta/2} \frac{k}{k+{\tilde c}}}\,.
\eq
\end{subequations}
These reproduce the results of earlier papers \cite{Yamaguchi1,Yamaguchi2,Revello}, which use different notations. Evidently they also generalize the ${\tilde c}=0$ solutions.

\begin{figure*}
    \centering
    \includegraphics[width=0.49\textwidth,keepaspectratio]{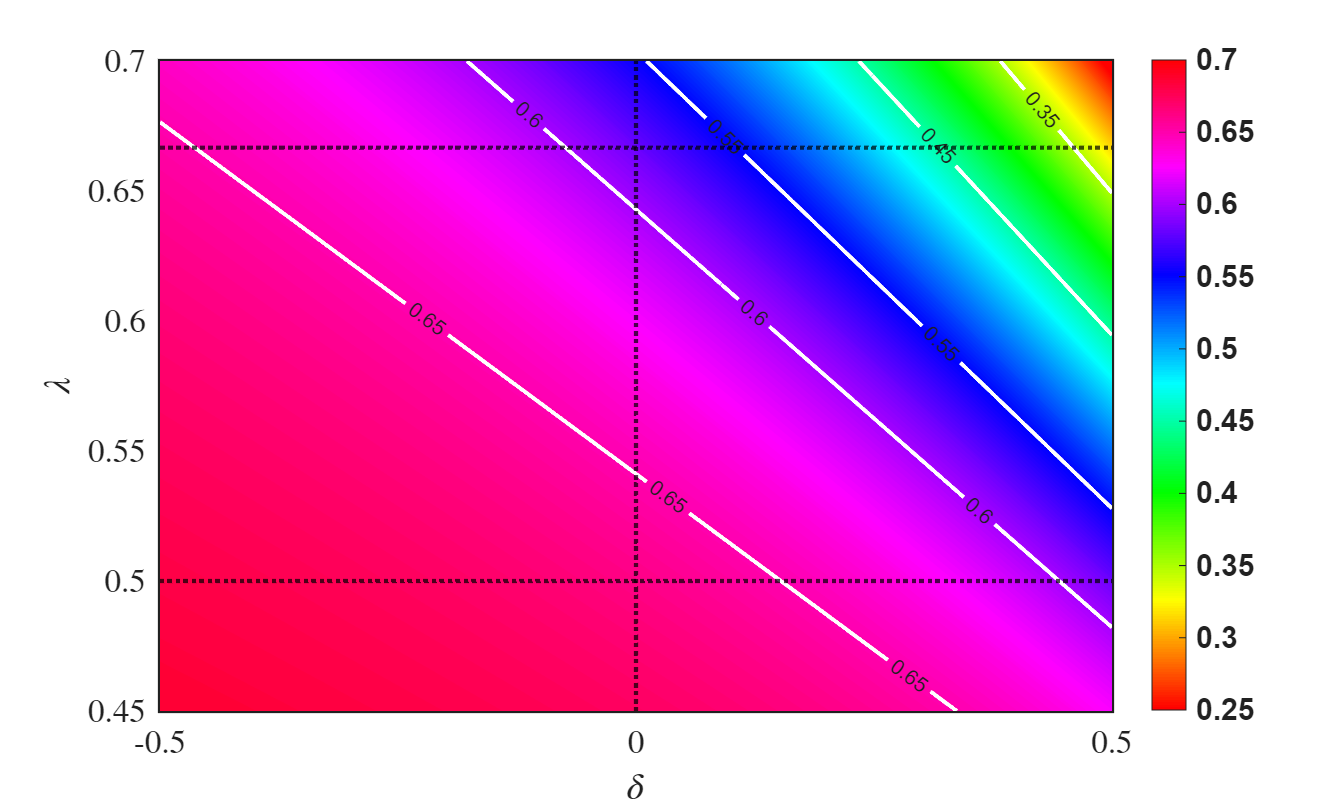}
    \includegraphics[width=0.49\textwidth,keepaspectratio]{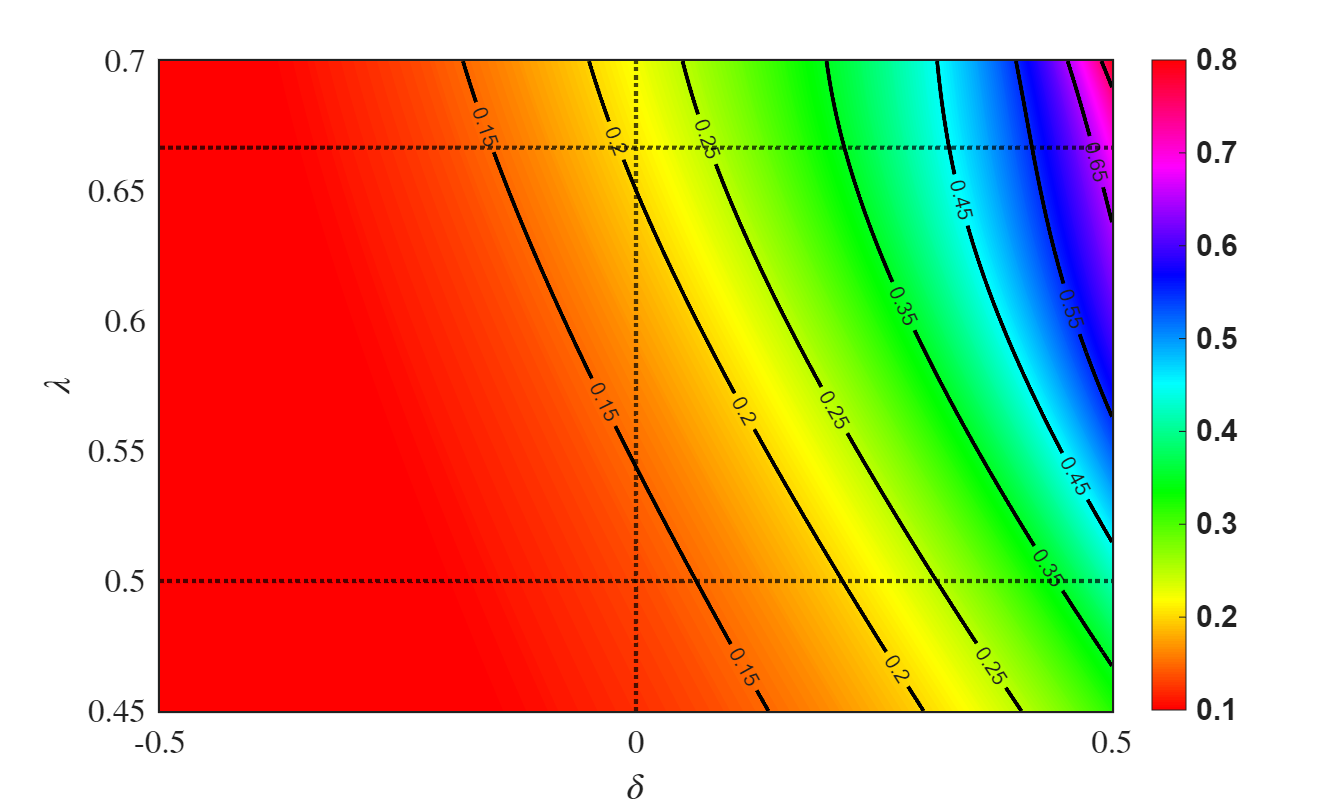}
    \includegraphics[width=0.49\textwidth,keepaspectratio]{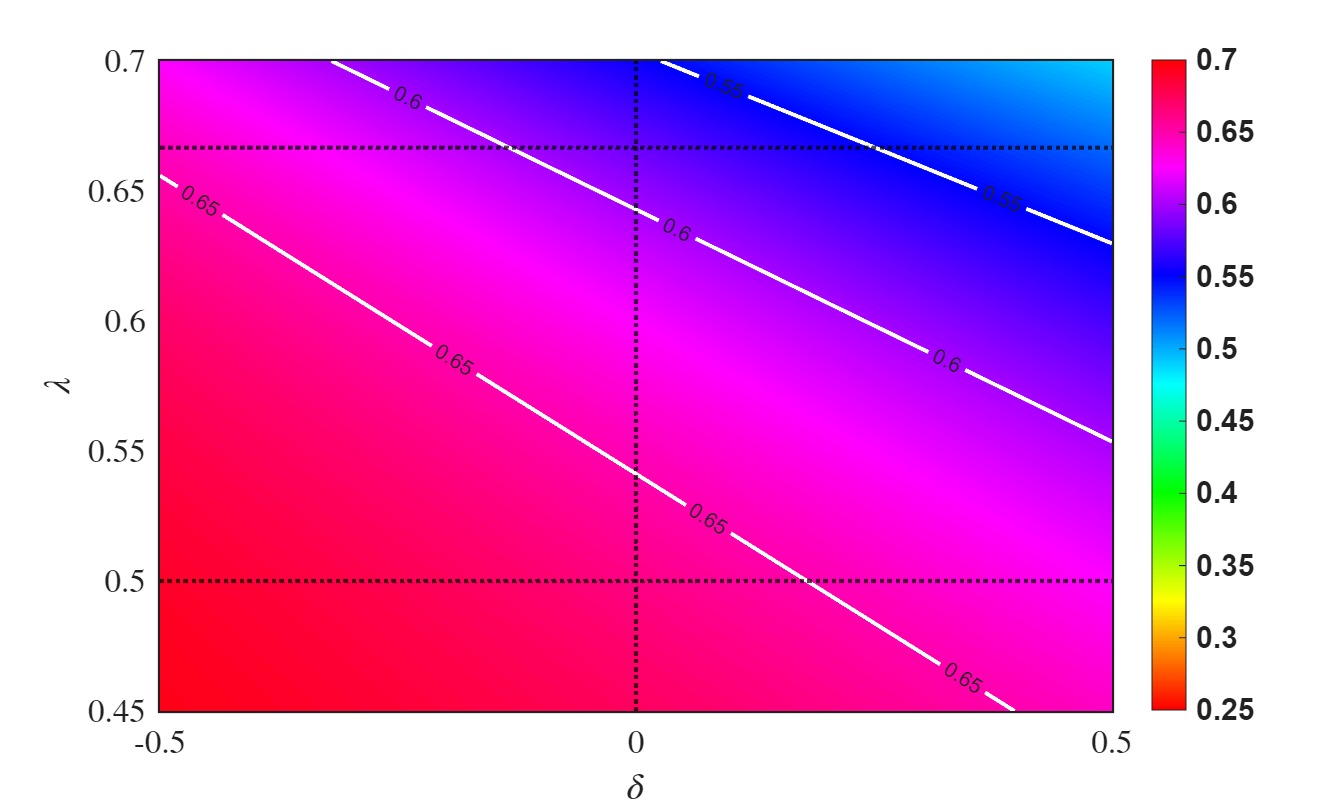}
    \includegraphics[width=0.49\textwidth,keepaspectratio]{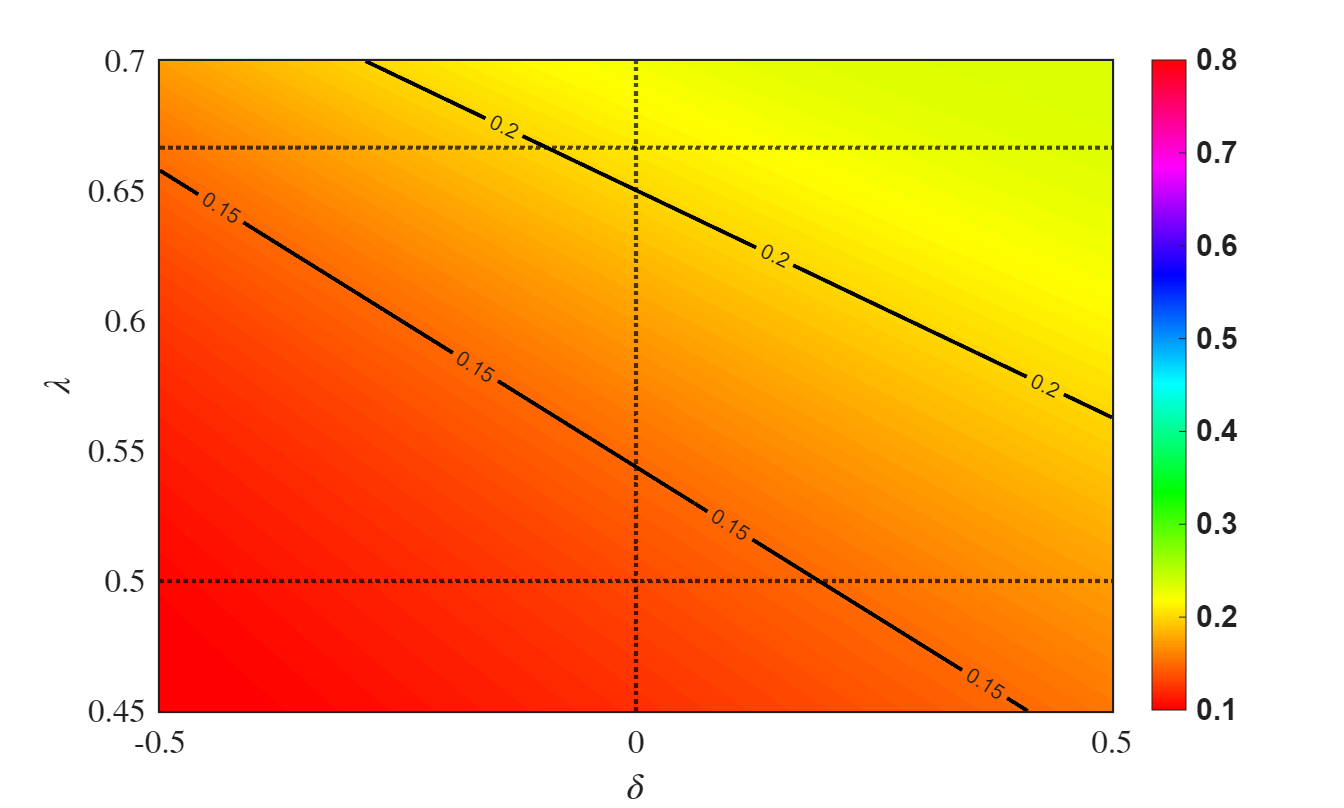}
    \caption{The linear scaling values of $v$ and $\epsilon$ (left and right panels respectively) for varying tension strings using the macroscopic and microscopic phenomenological approaches (top and bottom panels respectively), for relevant values of $\lambda$ and $\delta$, using $\tilde c=0.23$. The two horizontal dotted lines identify the radiation end matter eras ($\lambda=1/2$ and $\lambda=2/3$ respectively), while the vertical dotted line identifies the constant-tension case, $\delta=0$. To facilitate visual comparison, the colormaps for each plotted quantity are the same as those of Fig. \ref{fig1}. Some particular contour lines are also drawn for visual convenience.}
    \label{fig2}
\end{figure*}

Figure \ref{fig2} shows the corresponding linear scaling values of $v$ and $\epsilon$ for the two cases, for relevant values of $\lambda$ and $\delta$, and using the value of $\tilde c=0.23$ preferred by numerical simulations and the same form of $k(v)$. These plots make it clear that, for similar deviations from $\delta=0$, the macroscopic case has a much larger impact on the scaling values than the microscopic one.

In passing, we note that one may formally write a composite linear scaling solution
\begin{subequations}\label{composite}
\bq
L&=&\sqrt{\frac{k(k+{\tilde c})}{(2\lambda+\delta_2) (2-2\lambda-\delta_1)}}\, t\\
v&=&\sqrt{\frac{2-2\lambda-\delta_1}{2\lambda+\delta_2} \frac{k}{k+{\tilde c}}}\,,
\eq
\end{subequations}
where the macroscopic and microscopic cases correspond to $\delta_1\neq0$ and $\delta_2\neq0$ respectively. 

In the microscopic case, the choice of $\delta=-2\lambda$ (i.e., $\mu\propto a^{-2}$) would cancel the damping term, leading to a scaling solution matching that for a constant tension string network in Minkowski spacetime one, cf. Eq. (\ref{mink}). Moreover, if one thinks of these modifications, more broadly, as phenomenological changes of the stretching and damping terms, it is interesting to ask if some coordinated choices of $\delta_1$ and $\delta_2$ would leave the linear scaling standard solution, Eq. (\ref{linear}), unchanged. Apart from the trivial choice, $\delta_1=\delta_2=0$, there is indeed a second option, which depends on the expansion rate
\begin{subequations}
\bq
\delta_1&=&4(1-\lambda)\\
\delta_2&=&-4\lambda\,.
\eq
\end{subequations}
It follows that in the Minkowski spacetime limit ($\lambda=0$) one has $\delta_1=4$, $\delta_2=0$: there is no effective damping, but a considerable amount of stretching is needed. Conversely, in the inflation limit ($\lambda=1$) we require $\delta_1=0$, $\delta_2=-4$: no further stretching is needed, but one requires adequate negative damping to offset the fast expansion. Note that for all these cases one needs $\delta_1=\delta_2+4$, which we can interpret as the amount of effective stretching necessary to offset the effective damping.

Finally, let us briefly consider the special case of a universe fully dominated by cosmological constant, i.e. a de Sitter universe, in which
\begin{subequations}\label{lambda}
\bq
L&\propto&a\\
v&\propto&a^{-1}\,,
\eq
\end{subequations}
with $v_0L_0H=k$.
This solution is evidently unchanged in both the macroscopic and microscopic cases if we assume $\mu\propto t^\delta$ as we have been doing. On the other hand, if we assume $\mu\propto a^\delta$ the solutions do change. For the macroscopic case we have
\begin{subequations}\label{desitter}
\bq
L&\propto&a^{1+\delta/2}\\
v&\propto&a^{-1-\delta/2}\,,
\eq
\end{subequations}
provided $|\delta|<2$ and with $(1-\delta/2)v_0L_0H=k$. This also highlights the rescaling of the stretching term. On the other hand, the impact on damping is again much less significant, in this case due to the negligible velocities. Accordingly, in the microscopic case the solution given by Eq. (\ref{lambda}) still holds, although the consistency condition becomes $(1+\delta)v_0L_0H=k$

\section{\label{sect5}Interlude: contracting universes}

One can notice from Eqs. (\ref{rescales}--\ref{rescaled}) that a power-law time dependent string tension might effectively be seen as rescaling the Hubble parameter relevant for either the stretching or the damping term (without impacting the other term), $2\lambda\longrightarrow 2\lambda+\delta$. It follows that if $\delta<-2\lambda$ the sign of this rescaled term becomes negative, which would be akin to their behavior in a contracting universe. It should be kept in mind that this is only an imperfect analogy since, as we have seen, the rescaling applies to only one of the two lengths in the VOS dynamical equations. In this section we explore the case of contracting universes, which lead to ultra-relativistic networks (very briefly considered in \cite{Revello}), with the aim of understanding the extent to which the interpretation of the varying tension as the aforementioned partial rescaling is physically legitimate.

We begin by briefly recalling the dynamics of cosmic string networks in contracting universes \cite{Contracting0,Contracting}. When the Hubble parameter becomes negative the corresponding term in the velocity equation becomes an accelerating term (as opposed to a damping one) and therefore the string velocity will increase, become ultra-relativistic and gradually approach unity, with $\gamma v\propto a^{-2}$. This new acceleration term is offset by the fact that, in the curvature term, $k(v)$ also changes sign, becoming negative for $1/2<v^2<1$ \cite{VOS3}, but approaching zero as the velocity approaches unity. Physically, cosmic string networks in contracting universes behave as a radiation fluid, with $\rho\propto a^{-4}$. The network is conformally contracted, with the physical correlation length behaving as $L_{ph}\propto a$ while the corresponding invariant length behaving as $L_{inv}\propto a^2$. A more detailed discussion of these scaling solutions can be found in \cite{Contracting}.

In the sections of this work so far, the characteristic length $L$ was an invariant quantity, i.e. a measure of the invariant string energy, but in ultra-relativistic contexts it is clearer to deal with physical quantities. Bearing in mind that physical and invariant energies are related via $E_{inv}=\gamma E_{ph}$, where $\gamma$ is the Lorentz factor, one has $L_{ph}=\gamma^{1/2}L_{inv}$. Thus one can equivalently rewrite the VOS equations as
\begin{subequations}
\bq
2\frac{dL_{ph}}{dt}&=&\frac{L_{ph}}{\ell_s}+({\tilde c}_{ph}+k_{ph})v\\
\frac{d(\gamma v)}{dt}&=&\frac{\gamma k_{ph}}{L_{ph}}-\frac{\gamma v}{\ell_d}\,,
\eq
\end{subequations}
where we have also redefined ${\tilde c}_{ph}=\gamma^{1/2}{\tilde c}_{inv}$ and $k_{ph}=\gamma^{1/2}k_{inv}$. Note that in this form the damping length only appears in the velocity equation, while the stretching length remains in the characteristic length equation.

The extensions to the varying tension case are now clear. In the macroscopic case we have
\begin{subequations}
\bq
L_{ph}&\propto&\sqrt{\mu}a\\
(\gamma v)&\propto&a^{-2}\,,
\eq
\end{subequations}
while in the microscopic case
\begin{subequations}
\bq
L_{ph}&\propto&a\\
(\gamma v)&\propto&(\sqrt{\mu}a)^{-2}\,.
\eq
\end{subequations}
Thus only one or the other of the dynamical VOS model parameters is impacted. On the other hand, in both cases,
\begin{subequations}
\bq
L_{inv}&\propto&\sqrt{\mu}a^2\\
\rho&\propto&a^{-4}\,;
\eq
\end{subequations}
in other words, the radiation-like fluid behavior is maintained, as expected.

\section{\label{sect6}Realistic cosmological evolution}

Although studying cosmological scenarios with power law scale factors is mathematically simple and physically useful, these do not fully describe the evolution of our universe. In this section we discuss how a varying tension impacts the network properties during the transition from radiation to matter domination, and from matter domination to acceleration. For the standard constant tension case, this has been previously discussed in \cite{Azevedo:2017xev}, whose results we build upon and generalize.

\subsection{Radiation to matter}

For the radiation-to-matter transition we rely on the exact solution
\be
\frac{a(\tau)}{a_{eq}}=\left(\frac{\tau}{\tau_\star}\right)^2+2\left(\frac{\tau}{\tau_\star}\right)\,,
\ee
where $\tau$ is conformal time (related to physical time $t$ via $dt=ad\tau$), $\tau_\star=\tau_{eq}/(\sqrt{2}-1)$ and $a_{eq}$ and $\tau_{eq}$ are the scale factor and conformal time at the moment of equal radiation and matter densities respectively. It is therefore convenient to rewrite the VOS equations once more, this time as a function of conformal time, while simultaneously introducing a comoving version of the physical length scale, which we define as $\xi_{ph}=L_{ph}/a$. Then we can write
\begin{subequations}
\bq
2\frac{d\xi_{ph}}{d\tau}&=&({\tilde c}_{ph}+k_{ph})v+\frac{\mu'}{\mu}\xi_{ph}\\
\gamma^2\frac{dv}{d\tau}&=&\frac{k_{ph}}{\xi_{ph}}-\left(2\mathcal{H}+\frac{\mu'}{\mu}\right)v\,,
\eq
\end{subequations}
where $\mathcal{H}=a'/a$, primes denote derivatives with respect to conformal time, and for convenience we have already included the varying tension term in both the length scale and velocity equations; recall that each of these will apply only in the macroscopic and microscopic cases respectively.

\begin{figure*}
    \centering
    \includegraphics[width=0.49\textwidth,keepaspectratio]{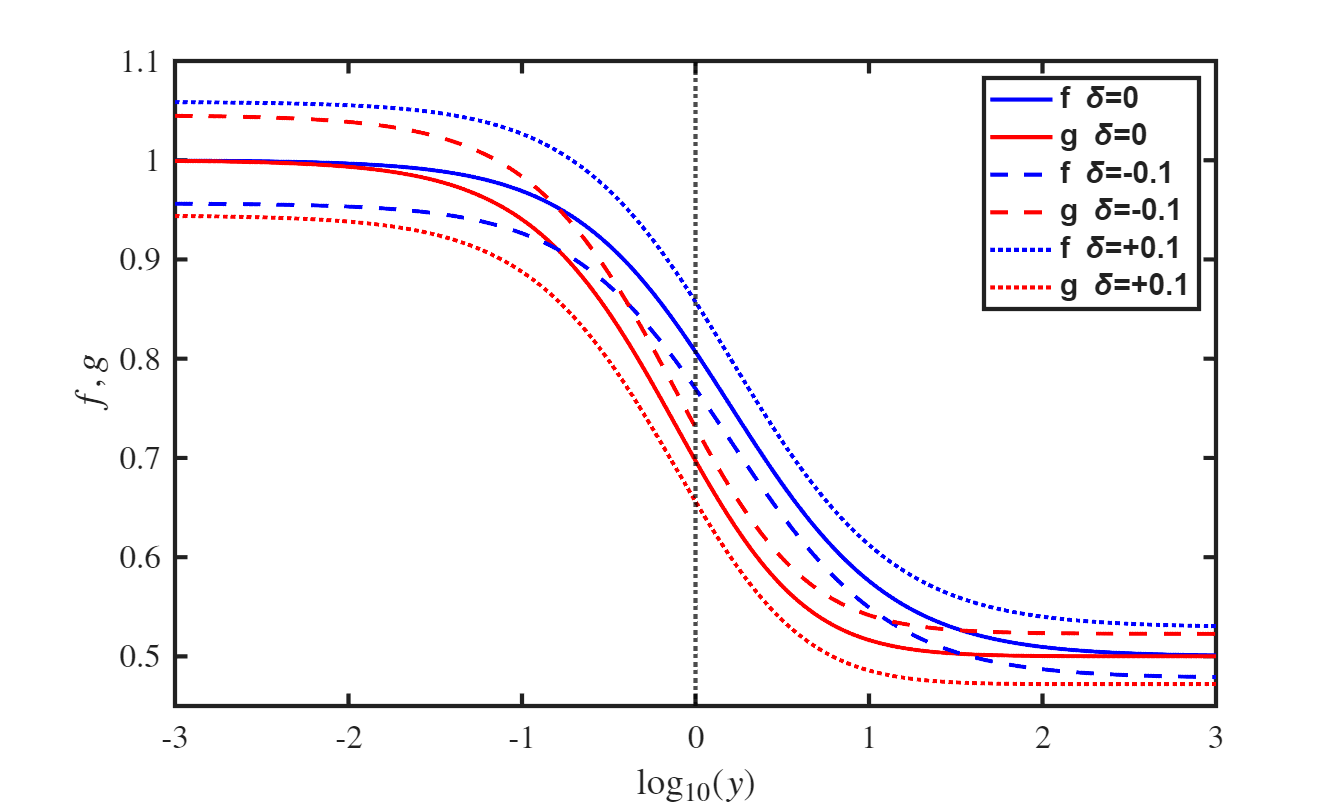}
    \includegraphics[width=0.49\textwidth,keepaspectratio]{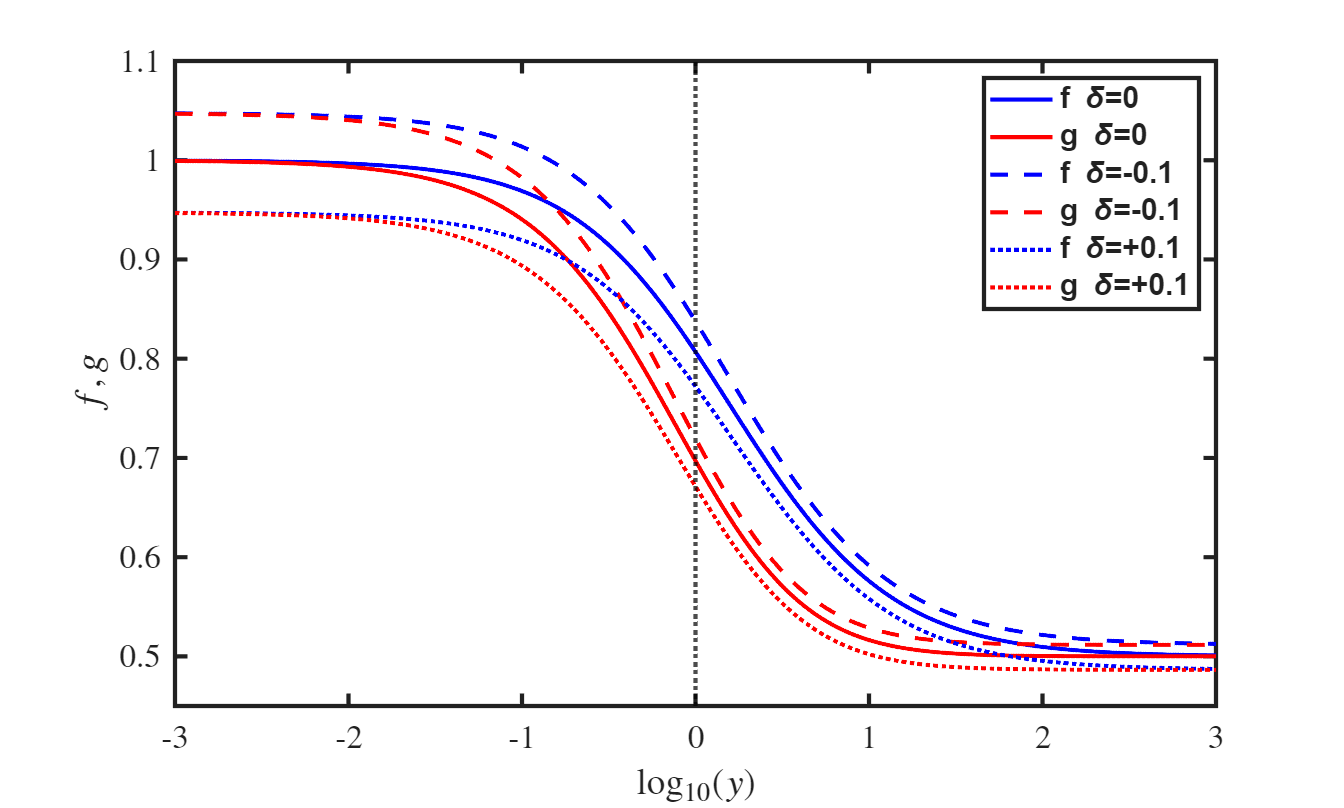}
    \caption{The behavior of transitions functions $f$ and $g$ (in blue and red, respectively) for $\delta=0$ (solid), $\delta=-1/10$ (dashed) and $\delta=+1/10$ (dotted). The left and right panels are for the macroscopic and microscopic cases respectively, and the black dotted line identifies the moment of equal radiation and matter densities.}
    \label{fig3}
\end{figure*}

In the standard case, noting that $a(t)\propto t^\lambda$ corresponds to $a(\tau)\propto \tau^{\lambda/1-\lambda}$ and defining $\zeta=\xi_{ph}/\tau$ we recover the conformal time version of the standard linear scaling solution
\begin{subequations}
\bq
\zeta^2&=&\frac{1}{4}k_{ph}(k_{ph}+{\tilde c_{ph}})\frac{1-\lambda}{\lambda}\\
v_0^2&=&\frac{k_{ph}}{k_{ph}+{\tilde c_{ph}}}\frac{1-\lambda}{\lambda}\,.
\eq
\end{subequations}
Note that the $\lambda$-dependent factor has a value of $1$ and $1/2$ in the radiation and matter eras respectively. For the evolution during the transition with a constant tension, \cite{Azevedo:2017xev} has shown that, using the fact that defect velocities are known (from both Nambu-Goto and field theory numerical simulations) to vary slowly and interpolate between the linear scaling solutions for the two epochs, one can write the following approximate solution for the network evolution across the transition
\begin{subequations}\label{functionsfg}
\bq
\zeta^2&=&\frac{1}{4}k_{ph}(k_{ph}+{\tilde c_{ph}})f(y)\\
v_0^2&=&\frac{k_{ph}}{k_{ph}+{\tilde c_{ph}}}g(y)\,,
\eq
\end{subequations}
where for convenience we have defined a dimensionless conformal time $y=\tau/\tau_\star$ and the two functions have the explicit form
\begin{subequations}\label{rmstandard}
\bq
f(y)&=&\frac{1}{2}+\frac{y-\ln{(1+y)}}{y^2}\\
g(y)&=&\frac{(2+y)^2}{4(1+y)^2f(y)}\,.
\eq
\end{subequations}
These are easily seen to have the correct asymptotic behavior in the deep radiation and matter eras, specifically
\begin{subequations}
\bq
f(0)=g(0)&=&1\\
f(\infty)=g(\infty)&=&\frac{1}{2}\,.
\eq
\end{subequations}
The solid red and blue lines of Fig. \ref{fig3} depict the standard behavior of both of these functions during the transition. It is important to note that although the timescale for the evolution of both functions is the same, the transition occurs first for $g$, i.e. for the velocities. Physically, the reason is clear: as the expansion rate increases the additional damping immediately impacts the velocities making them decrease; subsequently, as strings move more slowly the amount of energy losses due to loop production is also impacted.

Moving to the varying tension case, we keep the assumption $\mu\propto t^\delta$, which corresponds to $\mu(\tau)\propto \tau^{\delta'}$ (with $\delta'=\delta/(1-\delta)$) and for convenience start by writing the conformal time scaling solution analogous to that of Eq. (\ref{composite}), which has the form
\begin{subequations}
\bq
\zeta^2&=&\frac{k_{ph}(k_{ph}+{\tilde c_{ph}})}{\left(2-\delta'_1\right)\left(\frac{2\lambda}{1-\lambda}+\delta'_2\right)}\\
v_0^2&=&\frac{k_{ph}}{k_{ph}+{\tilde c_{ph}}}\frac{2-\delta'_1}{\frac{2\lambda}{1-\lambda}+\delta'_2}\,,
\eq
\end{subequations}
where as before $\delta'_1$ and $\delta'_2$ refer to the stretching (macroscopic) and damping (microscopic) corrections respectively---only one of them applying in each case. For the radiation-to-matter transition we can still expect transfer functions defined as in Eq. (\ref{functionsfg}). In the macroscopic case Eqs. (\ref{rmstandard}) become
\begin{subequations}
\begin{align}
f_1(y)&=\frac{{}_2F_1(1,1-\delta'_1;2-\delta'_1;-y)}{(\delta'_1-1)y}\notag \\
& - \frac{\delta'_1(y+1)-(y+2)}{(\delta'_1-2)(\delta'_1-1)y} \\
g_1(y)&=\frac{(2+y)^2}{4(1+y)^2f_1(y)}\,,
\end{align}
\end{subequations}
where ${}_2F_1$ is the Gaussian hypergeometric function. In this case the asymptotic values are
\begin{subequations}
\bq
f_1(0)=2f_1(\infty)&=&\frac{2}{2-\delta'_1}=\frac{2(1-\delta_1)}{2-3\delta_1}\\
g_1(0)=2g_1(\infty)&=&1-\frac{1}{2}\delta'_1=\frac{2-3\delta_1}{2(1-\delta_1)}\,.
\eq
\end{subequations}
Instead of using the hypergeometric function, one can also write $f_1$ in an equivalent but numerically more enlightening form
\be
f_1(y)=\frac{1}{y^{2-\delta'_1}}\int_0^y \frac{x^{1-\delta'_1}(2+x)}{1+x}dx\,.
\ee

On the other hand, in the microscopic case we find
\begin{subequations}
\begin{align}
f_2(y)&=\frac{(\delta'_2y/4+y+2)}{2(\delta'_2/4+1)^2y}\\
&-\frac{(\delta'_2/2+1)}{(\delta'_2/4+1)^3y^2}\ln{\left[\frac{\delta'_2(y+2)/4+y+1}{\delta'_2/2+1}\right]}\notag \\
g_2(y)&=\frac{4(2+y)^2}{[4(1+y)+\delta'_2(2+y)]^2f_2(y)}\,,
\end{align}
\end{subequations}
whose asymptotic values are
\begin{subequations}
\bq
f_2(0)=g_2(0)&=&\frac{2}{\delta'_2+2}=\frac{2(1-\delta_2)}{2-\delta_2}\\
f_2(\infty)=g_2(\infty)&=&\frac{2}{\delta'_2+4}=\frac{2(1-\delta_2)}{4-3\delta_2}\,.
\eq
\end{subequations}
In all the above cases one can check that these solutions reduce to the standard one, given by Eqs. (\ref{rmstandard}), when $\delta'_1=0$ and $\delta'_2=0$.

Figure \ref{fig3} contrasts the behavior of the $f$ and $g$ functions, for networks with slightly growing and decaying tensions, $\delta=\pm1/10$, in both the macroscopic and microscopic cases, and also shows the standard (constant tension) case to facilitate the comparison. One noteworthy common feature of is that it is always the case that the network velocities are impacted earlier in the transition than the characteristic length, but there are also two important differences between the two cases. The first and more straightforward mathematical one is that the hard constraint $\delta_1<2/3$ applies to the macroscopic case, while the analogous microscopic one is the weaker $\delta_2<2$. The second and physically more interesting one, is that we can again explicitly see the different physical impacts of phenomenologically changing the stretching or the damping terms. 

In the standard constant tension case the functions $f$ and $g$ have identical values in both asymptotic limits (the deep radiation era, $y\to0$, and the deep matter era, $y\to\infty$), even though they evolve differently between these limits, and moreover for each of the functions the ratio of its radiation to matter asymptotic values is the same and equals two. In the macroscopic case the latter of these still applies, but the former no longer holds: by effectively adding a stretching term in the characteristic length (or density) equation the impacts on the network's characteristic comoving length and velocities will be different, but the impact of this stretching addition is proportionally the same in both epochs (since otherwise no stretching-like term would exist in the $\xi_{ph}$ equation). In the microscopic case, we find the opposite situation: the two functions still have identical values in both asymptotic limits (since the rescaled damping impacts velocities directly and densities indirectly, as before), but the ratio of radiation to matter asymptotic values is no longer two. The reason for the latter is that in this case there is already some damping in both epochs, so adding the same amount of additional damping will have proportionally different impacts in both epochs, so the ratio must change. In this limited sense the microscopic approach is again a less drastic change, with respect to the constant tension, than the macroscopic one.

\subsection{Matter to acceleration}

The exploration of the network evolution during this more recent transition is, to some extent, analogous to the one in the previous subsection, with two differences which we will point out in what follows. As discussed in \cite{Azevedo:2017xev}, working in terms of physical on conformal time is not particularly convenient in this case; instead, one can work in terms of redshift or, almost equivalently, the scale factor. In this case the VOS equations read
\begin{subequations}
\begin{align}
2\frac{d\ln{\xi_{ph}}}{d\ln{a}}&=\frac{({\tilde c}_{ph}+k_{ph})v}{Ha\xi_{ph}}+\frac{d\ln{\mu}}{d\ln{a}} \\
\gamma^2\frac{dv}{d\ln{a}}&=\frac{k_{ph}}{Ha\xi_{ph}}-\left(2+\frac{d\ln{\mu}}{d\ln{a}}\right)v  \,,
\end{align}
\end{subequations}
where again, for convenience, we have already included the macroscopic and microscopic  running tension terms in the first and second equations respectively, bearing in mind that at most one of them will apply at a time.  We will assume a flat $\Lambda$CDM model, with a Friedmann equation
\be
H^2(z)=H^2_0\left[\Omega_m a^{-3}+\Omega_\Lambda\right]\,,
\ee
subject to $\Omega_m+\Omega_\Lambda=1$, and a string tension varying as $\mu(a)\propto a^\delta$. When plotting a specific example of this evolution, we will assume $\Omega_m=0.3$ in agreement with contemporary observations.

As in the radiation to matter transition case, one can proceed under the slow velocity change assumption, and look for solutions of the form
\begin{subequations}
\bq
H_0^2\xi^2_{ph}&=&\frac{1}{4}k_{ph}(k_{ph}+{\tilde c_{ph}})f(a)\\
v^2&=&\frac{k_{ph}}{k_{ph}+{\tilde c_{ph}}}g(a)\,,
\eq
\end{subequations}
although in this case there are no detailed Nambu-Goto or field theory simulations which can assess the validity of this assumption (both such simulations are technically challenging for vary fast expansion rates). Nevertheless, comparing the result of the numerical solution of the system of the two VOS equations with that obtained by the slow velocity change approximation shows that it is adequate for our purposes.

In the standard (constant tension) case, we have
\begin{subequations}
\begin{align}
f(a)&=\frac{1}{M}\ln{\left[\frac{\left[\Omega_m^{1/3}+\Omega_\Lambda^{1/3}a\right]^2}{\Omega_m^{2/3}-(\Omega_m\Omega_\Lambda)^{1/3}a+\Omega_\Lambda^{2/3}a^2} \right]}\notag\\
&+\frac{2\sqrt{3}}{M}\arctan{\left(\frac{2\Omega_\Lambda^{1/3}a-\Omega_m^{1/3}}{\sqrt{3}\Omega_m^{1/3}}\right)} + \frac{\pi}{\sqrt{3}M} \\
g(a)&=\frac{a}{\left[\Omega_m+\Omega_\Lambda a^3\right]f(a)}\,,
\end{align}
\end{subequations}
where for convenience we have defined the constant $M=3\Omega_m^{2/3}\Omega_\Lambda^{1/3}$. Bearing in mind the relations between scale factor and time, one can show that in the limit $a\longrightarrow0$ one recovers the matter-era linear scaling solution (in this limit $f(a)\propto a\propto \tau^2$), while in the opposite limit $a\longrightarrow\infty$, corresponding to the far future, one obtains the de Sitter solution discussed in Eq. (\ref{lambda}).

\begin{figure*}
    \centering
    \includegraphics[width=0.49\textwidth,keepaspectratio]{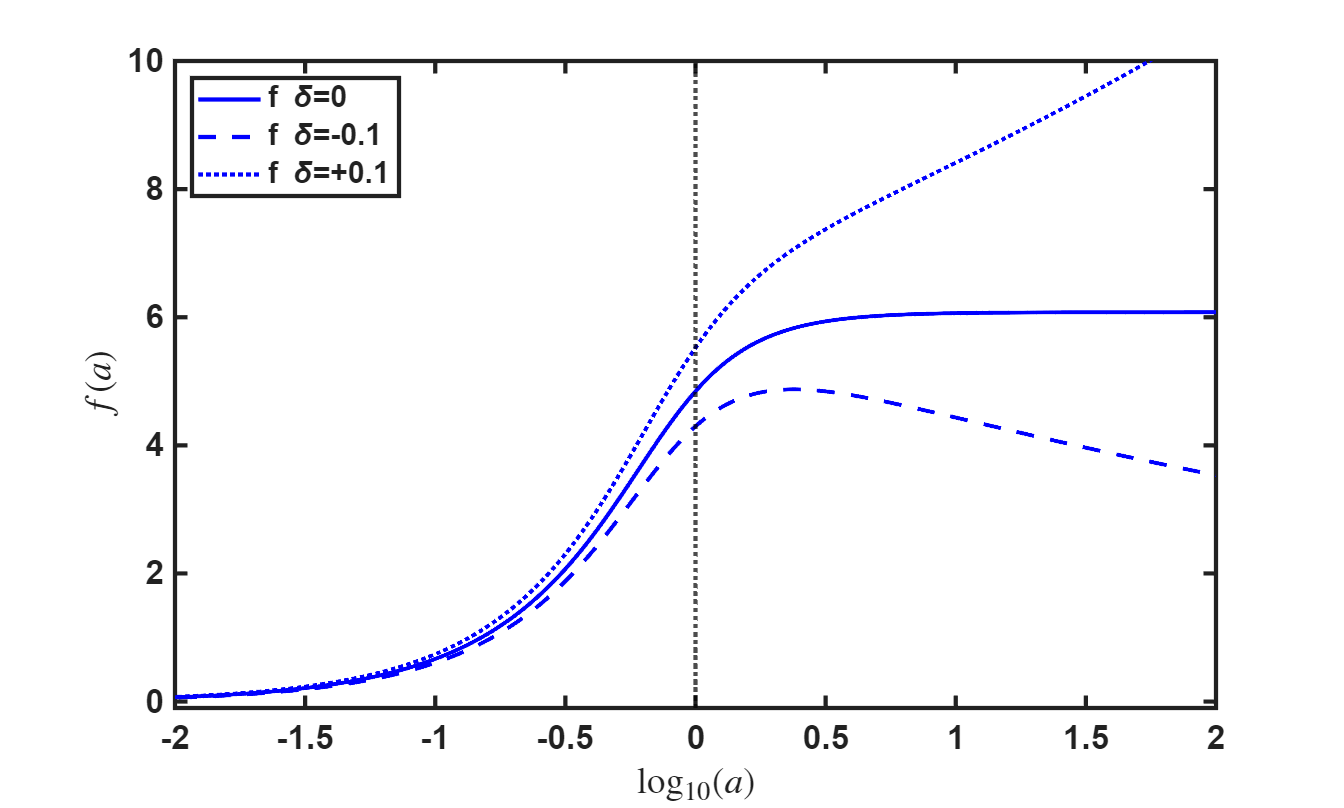}
    \includegraphics[width=0.49\textwidth,keepaspectratio]{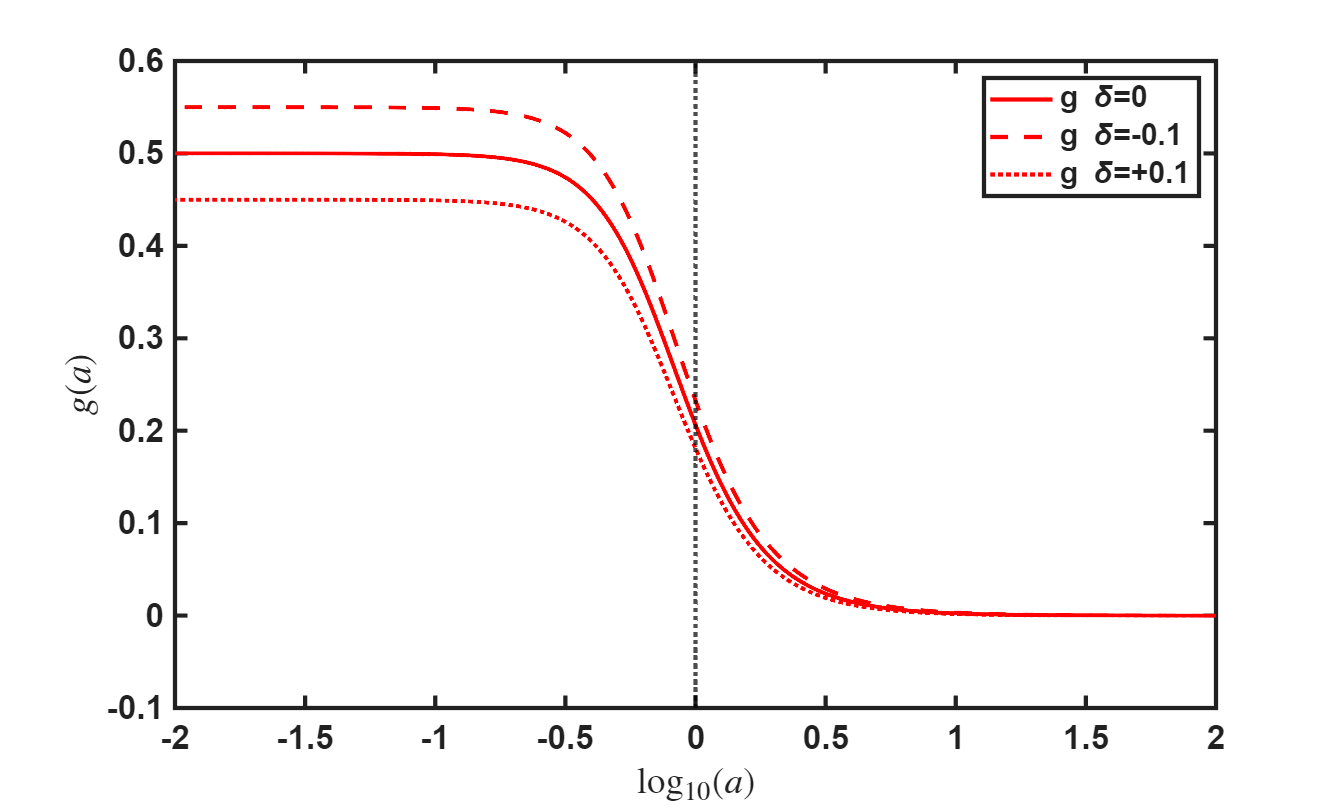}
    \includegraphics[width=0.49\textwidth,keepaspectratio]{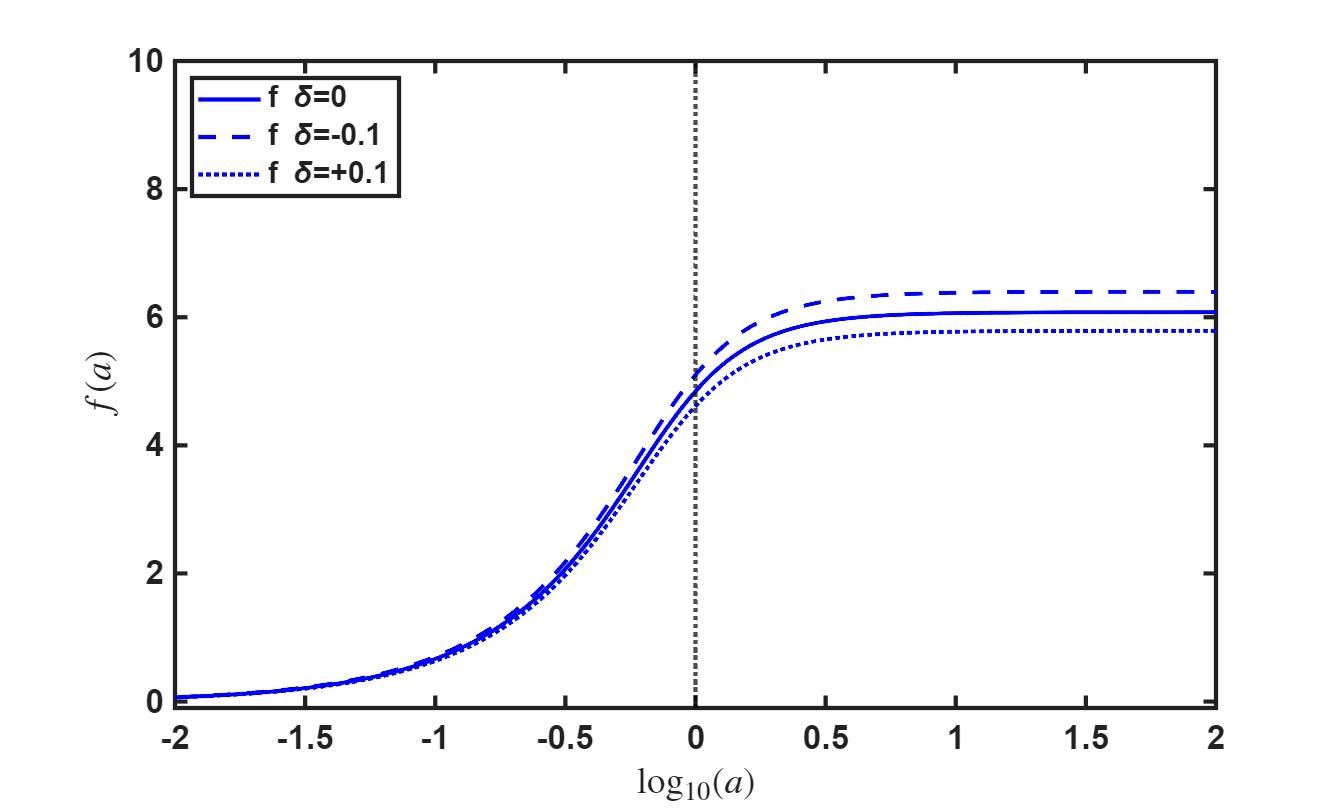}
    \includegraphics[width=0.49\textwidth,keepaspectratio]{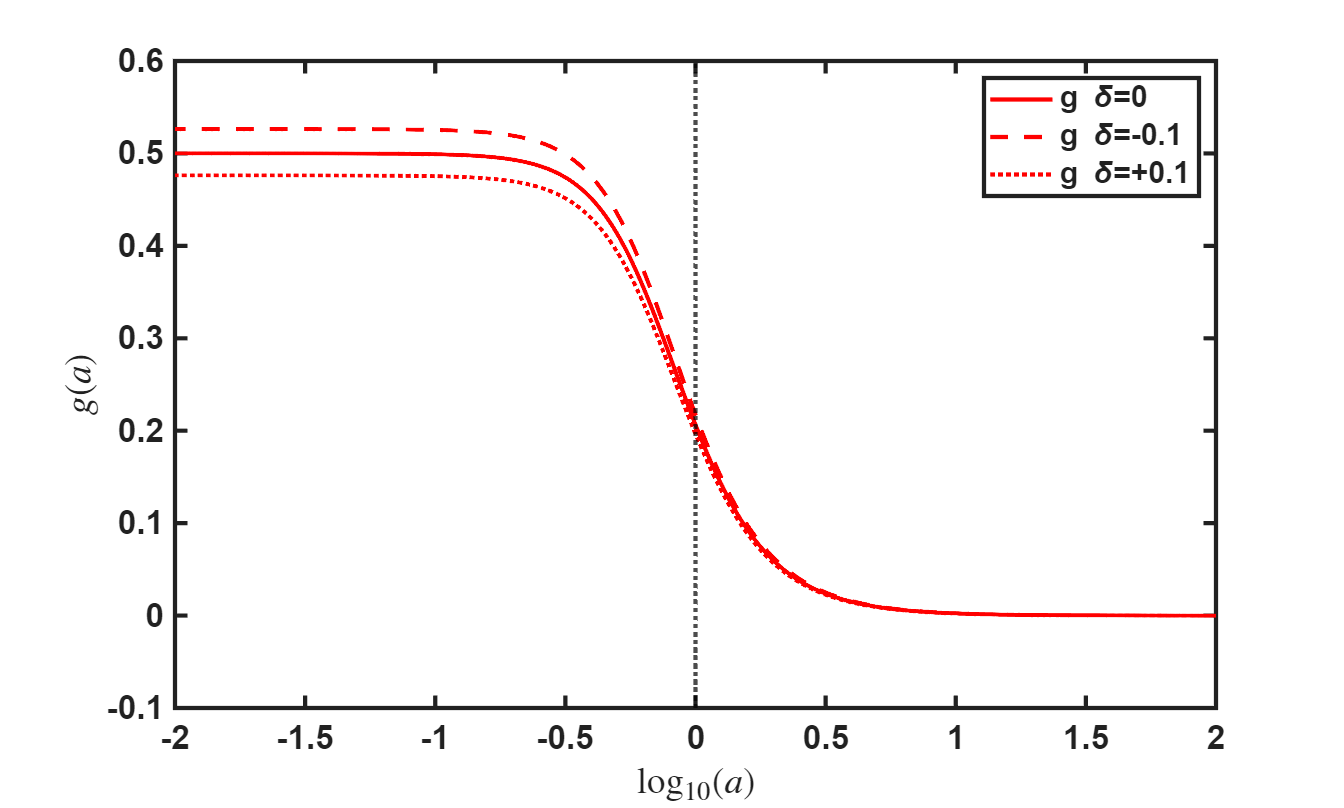}
    \caption{The behavior of transition functions $f$ and $g$ (in blue and red, left and right side panels respectively) for $\delta=0$ (solid), $\delta=-1/10$ (dashed) and $\delta=+1/10$ (dotted). The top and bottom panels are for the macroscopic and microscopic cases respectively, and the vertical dotted line identifies the present day.}
    \label{fig4}
\end{figure*}
\begin{figure*}
    \centering
    \includegraphics[width=0.49\textwidth,keepaspectratio]{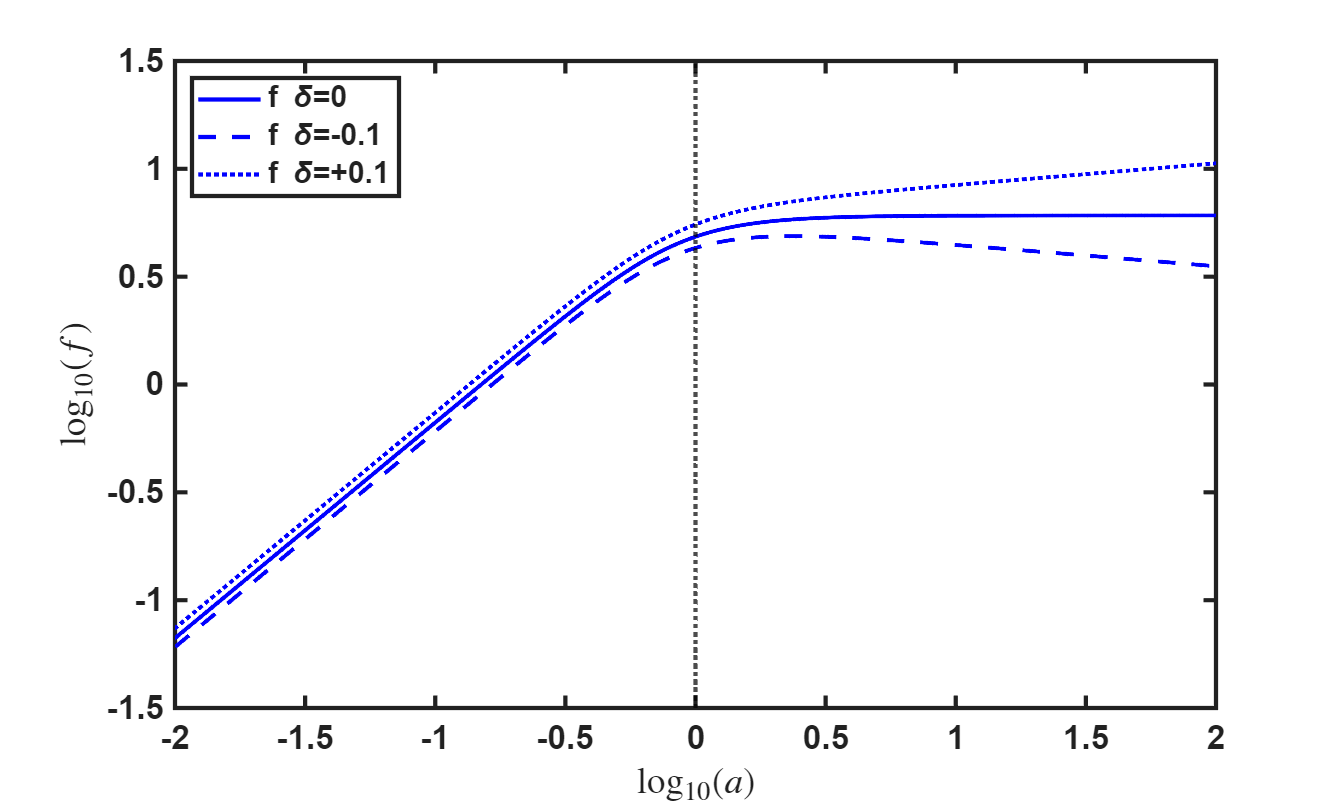}
    \includegraphics[width=0.49\textwidth,keepaspectratio]{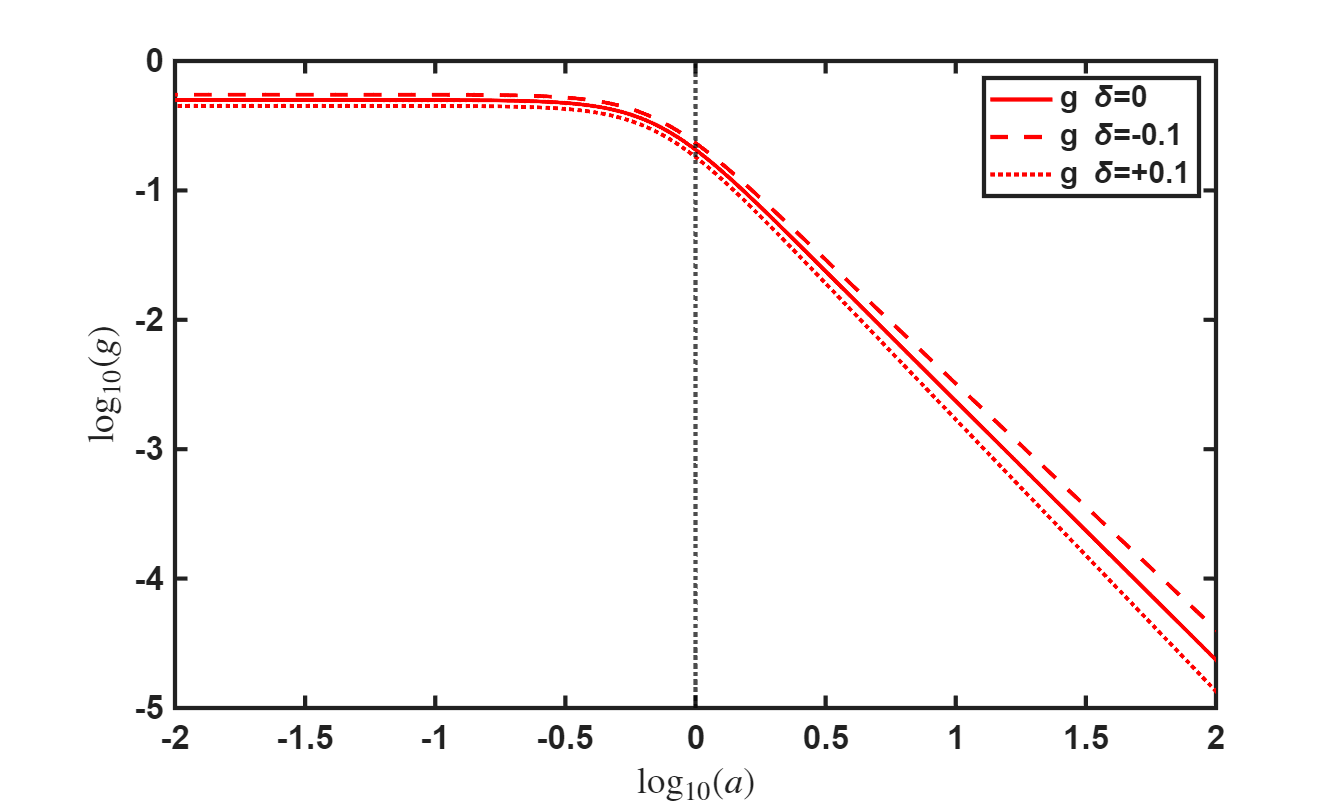}
    \includegraphics[width=0.49\textwidth,keepaspectratio]{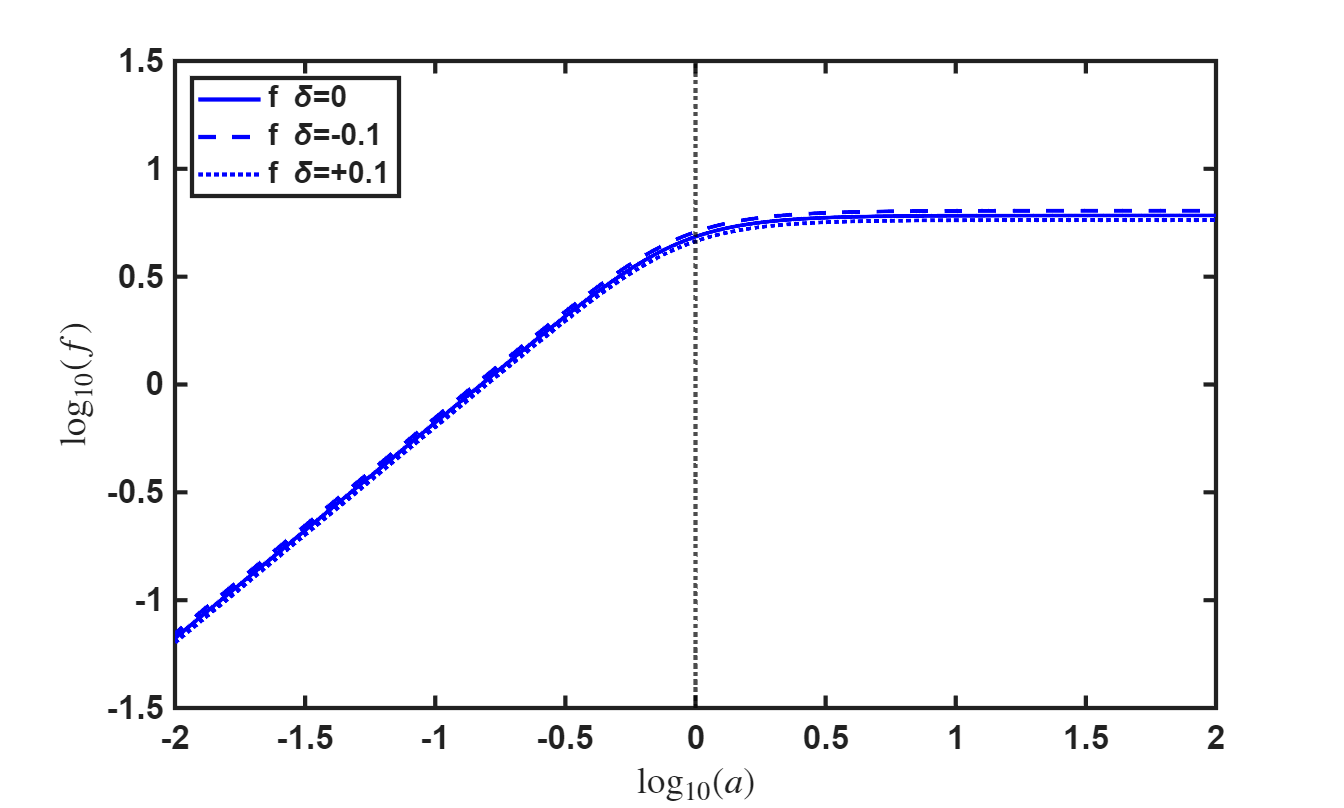}
    \includegraphics[width=0.49\textwidth,keepaspectratio]{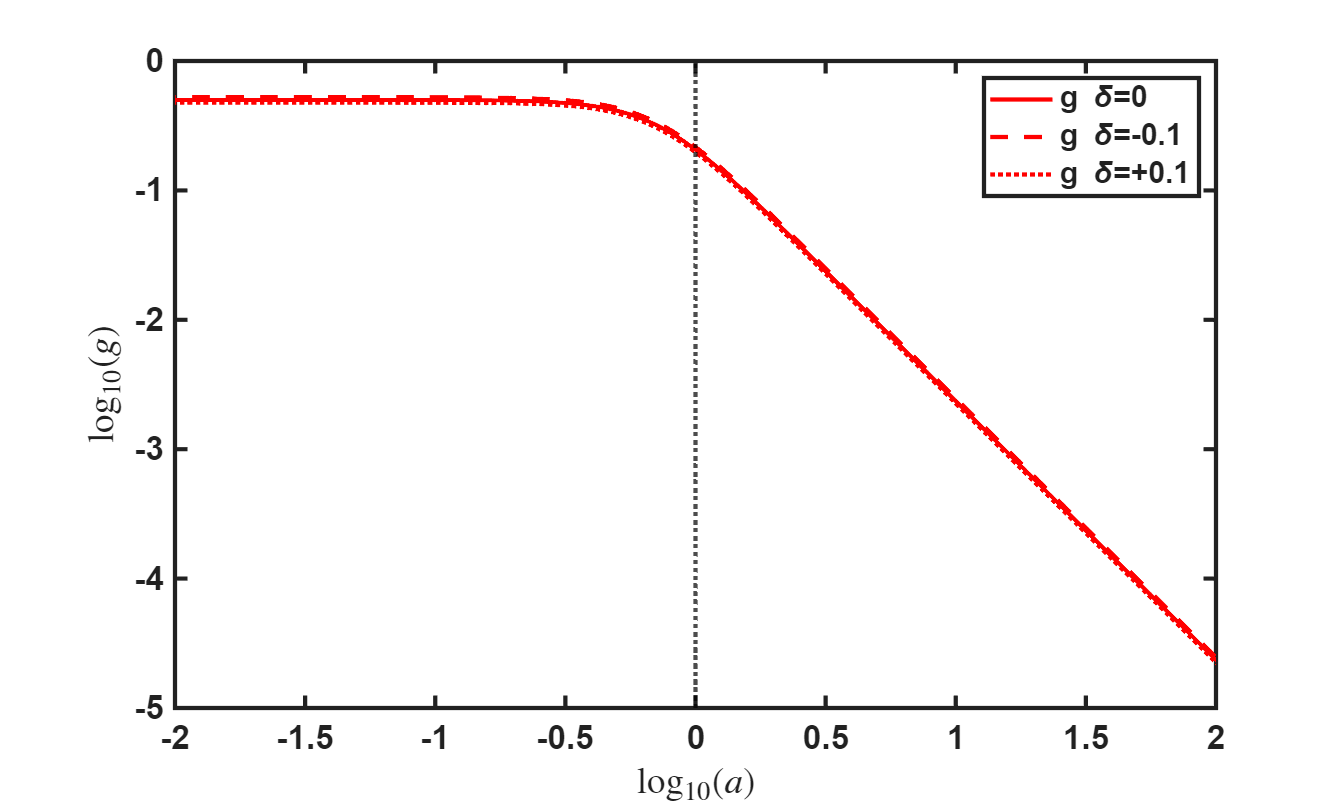}
    \caption{Same as Fig. \ref{fig4}, with the vertical axes in logarithmic rather than linear  scale.}
    \label{fig5}
\end{figure*}

Now, in the macroscopic case, the previous solution generalizes to
\begin{subequations}
\begin{align}
f_1(a)&=2\frac{{}_2F_1(1,(1-\delta_1)/3;(4-\delta_1)/3;-\Omega_\Lambda a^3/\Omega_m)}{(1-\delta_1)\Omega_m} a \\
g_1(a)&=\frac{a}{\left[\Omega_m+\Omega_\Lambda a^3\right]f_1(a)}\,,
\end{align}
\end{subequations}
where again $f_1$ can alternatively be written
\be
f_1(a)=2a^{\delta_1}\int_0^a \frac{x^{-\delta_1}}{\Omega_m+\Omega_\Lambda x^3}dx\,,
\ee
while in the microscopic case the solution can be written in the simpler form
\begin{subequations}
\begin{align}
f_2(a)&=\frac{f(a)}{1+\delta_2/2}\\
g_2(a)&=\frac{g(a)}{1+\delta_2/2}\,.
\end{align}
\end{subequations}
One can again check that the various appropriate limits are recovered. Figures \ref{fig4} and \ref{fig5} again contrast the behavior of the $f$ and $g$ functions, for networks with $\delta=\pm1/10$, in both the macroscopic and microscopic cases with standard (constant tension) case. In particular, the logarithmic scale plots serve to confirm the previously derived results in the matter era and inflation limits; the solution reported in Eq. (\ref{desitter}) is particularly salient in the logarithmic plots.

\section{\label{sect7}Conclusions}

We have provided a detailed exploration of the consequences, specifically in terms of scaling and other solutions, of two phenomenological assumptions for describing varying tension cosmic strings, in cosmological and non-cosmological settings, using the VOS model. The two differ on the level at which the variation, which is stipulate to only depend on time (without spatial variations) is assumed to occur, and we have therefore dubbed them the macroscopic and microscopic assumptions. They can both be seen as approximations to the more realistic case of wiggly cosmic strings, but are interesting phenomenologically because the two correspond to opposite assumptions on the impact of the varying tension: in one approach one rescales the damping term in the network's dynamical equations without affecting the stretching term, while in the other one does the opposite.

\begin{table}
    \centering
	\begin{tabular}{|c|cc|} 
		\hline
		Scaling solution & Macroscopic &  Microscopic \\
	\hline
    $\ell_f=const$ & Normalization & Unchanged \\
    $T=const$ & Scaling & Scaling \\
    Minkowski & Normalization & Scaling \\
    \hline
	Stretching & Scaling & Scaling \\
    Kibble & Scaling & Scaling \\
    Linear & Normalization & Normalization \\
	\hline
    de Sitter & Scaling & Normalization \\
    Contracting & Scaling & Scaling \\
    \hline
	\end{tabular}
	\caption{Impact of the varying tension on various scaling solutions studied in this work, in the macroscopic and microscopic phenomenological descriptions.}
	\label{tab1}
\end{table}

We have quantified how the above modeling assumptions, together with the assumption of power-law dependencies for the string tension, impact each of the standard scaling solutions, and also discussed the behavior of the network during the cosmologically realistic radiation-to-matter and the matter-to-acceleration transitions. Depending on the physical context the time dependence of the scaling laws (for the characteristic length, the rms velocity, or both) can be affected, or the impact of the varying tension might be restricted to changing the quantitative normalization of these laws without affecting their time dependence; these different behaviors are summarized in Table \ref{tab1}. The best known linear (scale invariant) scaling solution is an example of the latter behavior. As for the network density, given that $\rho=\mu/L^2$ one might generically expect it to be impacted, but this is not necessarily the case: as we have shown, in several contexts the varying tension effect is canceled out by a corresponding change in the behavior of the characteristic length. Overall, our results show that for the same amount of tension variation, a change in the stretching length scale (implied by the macroscopic assumption) tends to have a more significant impact on the network than a change in the damping length (implied by the microscopic assumption).

Our analysis focused on the impact of a varying tension on the long string network; an analogous study could be done for individual loops, whose length evolution is given by Eq. (\ref{loopeqn}). In this regard, we note that it is in principle possible, in some of the parameter space (for sufficiently fast decaying tensions), that loop lengths grow in comoving coordinates \cite{Conlon,Revello}. In this case the loops can have a significant probability of reconnecting to the long string network, unlike in the standard scenario where this probability is known to be negligible. In the context of the VOS model, this possibility can phenomenologically be accounted for with an 'effective' loop chopping efficiency ${\tilde c}$, with a smaller than standard value---which might even be negative, in extreme (though arguably unphysical) situations. Quantifying how much smaller this parameter should be, for a specific decaying tension rate, requires dedicated numerical simulations.

Looking ahead, our results lead to at least three further lines of inquiry. It should be possible to confirm these solutions both in Nambu-Goto and in field theory simulations, although in the latter case one must be careful to simulate physical networks; the commonly used Press-Ryden-Spergel algorithm \cite{PRS} would not be suitable for this purpose since it can itself be interpreted as a peculiar type of varying tension, and therefore one should expect a degeneracy between the two effects. On the modeling side, a more thorough discussion of the relation between the two phenomenological assumptions and the formally more robust case of wiggly strings would be interesting; in that case the scaling laws are somewhat different \cite{Almeida_2021,Almeida_2022}, but there are still some broad similarities. Along the same lines, studying cases where the string tension is not assumed to evolve as a power law (of time or the scale factor, as the case may be) is also warranted. One such example is the case of cosmological axion strings, in which one effectively has a logarithmic dependence, which may or may not propagate into the network's scaling properties \cite{Axion1,Axion2}. These topics will be the subject of subsequent work.


\begin{acknowledgments}
This work was financed by Portuguese funds through FCT (Funda\c c\~ao para a Ci\^encia e a Tecnologia) in the framework of the project 2022.04048.PTDC (Phi in the Sky, DOI 10.54499/2022.04048.PTDC). CJM also acknowledges FCT and POCH/FSE (EC) support through Investigador FCT Contract 2021.01214.CEECIND/CP1658/CT0001 (DOI 10.54499/2021.01214.CEECIND/CP1658/CT0001). 
\end{acknowledgments}
\bibliography{article}
\end{document}